\documentclass[nojss]{jss}

\usepackage[utf8]{inputenc}
\graphicspath{{Figures/}}
\DeclareGraphicsExtensions{.eps,. ps,. pdf, .jpg, .png}

\usepackage{orcidlink}
\usepackage{subfigure}
\usepackage{float}
\usepackage{bm}
\usepackage{amsmath, amsfonts}
\usepackage{setspace}
\usepackage{booktabs}
\usepackage{dsfont}
\usepackage{accents}
\usepackage{mathtools}

\usepackage{pifont}
\newcommand{\cmark}{\ding{51}}%
\newcommand{\xmark}{\ding{55}}%
\newcommand{\bcd}{\boldsymbol{\cdot}}%

\newcommand{\rY}{Y}
\newcommand{\rX}{\mX}
\newcommand{\rz}{z}
\newcommand{\ry}{y}
\newcommand{\rx}{\xvec}
\newcommand{\pYx}{F_{\rY | \rX = \rx}}
\newcommand{\pZ}{F_Z}
\newcommand{\dZ}{f_Z}
\newcommand{\h}{h}
\newcommand{\hY}{h_\rY}
\newcommand{\basisy}{\avec}
\newcommand{\bern}[1]{\avec_{\text{Bs},#1}}
\newcommand{\basisx}{\bvec}
\newcommand{\eparm}{\vartheta}
\newcommand{\parm}{\varthetavec}

\newcommand{\shiftparm}{\betavec}

\newcommand{\ie}{{i.e.,}~}
\newcommand{\eg}{{e.g.,}~}
\newcommand{\cf}{{cf.}~}

\newcommand{\RR}{\mathbb{R}}

\newcommand{\given}{\mid}

 \DeclareMathOperator{\expit}{expit}

\def \avec {\text{\boldmath$a$}}
\def \bvec {\text{\boldmath$b$}}
\def \evec {\text{\boldmath$e$}}
\def \svec {\text{\boldmath$s$}}
\def \xvec {\text{\boldmath$x$}}

\def \rX {\text{\boldmath$X$}}

 \def \calF {\mathcal F}
 \def \calX {\mathcal X}
 \def \calY {\mathcal Y}

\def \betavec         {\text{\boldmath$\beta$}}
\def \varthetavec     {\text{\boldmath$\vartheta$}}
\def \phivec          {\text{\boldmath$\phi$}}
\def \psivec          {\text{\boldmath$\psi$}}
\def \omegavec        {\text{\boldmath$\omega$}}

\newcommand{\ubar}[1]{\underaccent{\bar}{#1}}

\author{%
  Lucas Kook \\ Vienna University of \\
  Economics and Business \And
Philipp F. M. Baumann \\ ETH Zurich \AND
Oliver D\"urr \\ HTWG Konstanz \And
Beate Sick \\ University of Zurich \\ Zurich University of \\ Applied Sciences
\And
David R\"ugamer \\ LMU Munich \\ Munich Center
for \\ Machine Learning}

\newtheorem{example}{Example}

\newcommand{\deeptrafo}{deeptrafo}
\newcommand{\dctm}{DCTM}
\newcommand{\msim}{$\mathtt{\sim}$~}
\newcommand{\lodm}{\code{list\_of\_deep\_models}}
\title{Estimating Conditional Distributions with Neural Networks 
using \proglang{R}~Package \pkg{\deeptrafo}}

\Plainauthor{Kook, Baumann, D\"urr, Sick and R\"ugamer}    
\Plaintitle{\deeptrafo: Estimating Conditional Distributions 
with Neural Networks in R} 
\Shorttitle{\pkg{\deeptrafo}: Estimating Conditional Distributions 
with Neural Networks in \proglang{R}} 

\Abstract{%
Contemporary empirical applications frequently require flexible regression
models for complex response types and large tabular or non-tabular, including
image or text, data. Classical regression models either break down under the
computational load of processing such data or require additional manual feature
extraction to make these problems tractable. Here, we present \pkg{\deeptrafo},
a package for fitting flexible regression models for conditional distributions
using a \pkg{tensorflow} backend with numerous additional processors, such as
neural networks, penalties, and smoothing splines. Package \pkg{\deeptrafo}
implements deep conditional transformation models (\dctm{s}) for binary,
ordinal, count, survival, continuous, and time series responses, potentially
with uninformative censoring. Unlike other available methods, \dctm{s} do not
assume a parametric family of distributions for the response. Further, the data
analyst may trade off interpretability and flexibility by supplying custom
neural network architectures and smoothers for each term in an intuitive formula
interface. We demonstrate how to set up, fit, and work with \dctm{s} for several
response types. We further showcase how to construct ensembles of these models,
evaluate models using inbuilt cross-validation, and use other convenience
functions for \dctm{s} in several applications. Lastly, we discuss \dctm{s} in
light of other approaches to regression with non-tabular data.
}
\Plainkeywords{deep learning; distributional regression; neural networks;
transformation models}

\Address{
  Lucas Kook\\
  Institute for Statistics and Mathematics\\
  Vienna University of Economics and Business\\
  Welthandelsplatz 1, 1020 Vienna, Austria\\
  E-mail: \email{lucas.kook@wu.ac.at}
  
  David R{\"u}gamer\\
  Working Group Data Science\\
  Department of Statistics\\
  LMU Munich\\
  80539, Munich, Germany\\
  E-mail: \email{david.ruegamer@stat.uni-muenchen.de}
}

\newcommand{\cls}[1]{`\texttt{#1}'}
\DeclareMathOperator{\NLL}{NLL}

\begin{document}

\section{Introduction} \label{sec:intro}

Regression analysis aims to characterize the conditional distribution of a
response $\rY$ given a set of covariates $\rX$, thereby describing how changes
in the covariates propagate to the conditional distribution of $\rY$ given $\rX$
\citep{FahKneLanMar2013}. In this paper, we present \pkg{\deeptrafo}
\citep{pkg:deeptrafo}, an \proglang{R} package for estimating a broad class of
distributional regression models for various types of responses (continuous,
survival, count, ordinal, binary) using tabular or non-tabular (\eg image or
text) data or both. Package \pkg{\deeptrafo} is available from the Comprehensive
\proglang{R} Archive Network (CRAN) at
\url{https://CRAN.R-project.org/package=deeptrafo}. Due to a flexible
\pkg{tensorflow} \citep{pkg:tensorflow} backend and mini-batch optimization,
\pkg{\deeptrafo} not only scales well with non-tabular (imaging, text) data but
also big tabular data sets. Many well-known models fall into the class of
transformation models (TMs), such as normal linear regression (Lm), Cox
proportional hazards models (CoxPH), and proportional odds logistic regression
\citep[Polr,][]{hothorn2018most}. In the following, we review existing software
for fitting these models.

\paragraph{Existing software packages}
TMs for tabular data are implemented in \pkg{tram} \citep{pkg:tram} using
\pkg{mlt} \citep{pkg:mlt} and fitted via maximum likelihood, relying on
\pkg{alabama} \citep{pkg:alabama} and \pkg{BB} \citep{pkg:BB} for optimization.
Package \pkg{tram} provides an intuitive interface for fitting a multitude of
distributional regression models, ranging from shift and shift-scale
\citep{siegfried2022distribution} to tensor-product (or ``conditional'')
transformation models \citep{hothorn2014conditional}. Several extensions of
transformation models exist. For instance, \pkg{cotram} for count TMs
\citep{pkg:cotram}, \pkg{tramME} for mixed effects TMs and TMs including
smoothing splines \citep{pkg:tramME}, and \pkg{tramnet} as well as \pkg{tramvs}
for regularized TMs \citep{pkg:tramnet,pkg:tramvs}. Transformation boosting
machines \citep{pkg:tbm} and transformation trees and random forests
\citep{pkg:trtf} offer extensions to classical machine learning models.
Table~\ref{tab:tram} summarizes the commonalities and differences between the
packages implementing different (extensions of) transformation models in terms
of model classes, support for \pkg{mgcv}-based splines and
\pkg{tensorflow}-based neural networks and scalable optimization (via mini-batch
training, see Appendix~\ref{app:fac}). The \pkg{\deeptrafo} package is currently
the only package implementing transformation models which supports neural
network architectures enabling direct handling of text, image, and other deep
learning-related data without requiring feature engineering.
\begin{table}[!t]
\centering
\resizebox{\textwidth}{!}{%
\begin{tabular}{lrrrrr}
\toprule
\bf Package & \bf Model class & \bf Non-linear & \bf Splines &
\bf Neural networks & \bf Scalable optimization \\
\midrule
  \pkg{tram} & Linear TMs & \xmark & \xmark & \xmark & \xmark \\
  \pkg{cotram} & Count TMs & \xmark & \xmark & \xmark & \xmark \\
  \pkg{tramnet} & $L_1$/$L_2$-penalized TMs & \xmark & \xmark & \xmark & \xmark \\
  \pkg{tramvs} & $L_0$-penalized TMs & \xmark & \xmark & \xmark & \xmark \\
  \pkg{tbm} & Additive TMs & \cmark & \xmark & \xmark & \xmark \\
  \pkg{trtf} & Transformation forests & \cmark & \xmark & \xmark & \xmark \\
  \pkg{tramME} & Additive mixed TMs & \cmark & \cmark & \xmark & \xmark \\
  \pkg{deeptrafo} & Additive TMs & \cmark & \cmark & \cmark & \cmark \\
\bottomrule
\end{tabular}}
\caption{%
Overview of packages for estimating different classes of transformation models.
Packages \pkg{tramME}, \pkg{tbm}, \pkg{trtf}, and \pkg{deeptrafo} support 
estimation of non-linear TMs. Specifically, \pkg{tramME} supports splines from
\pkg{mgcv}, \pkg{tbm} fits non-linear model components via score-based boosting,
\pkg{trtf} fits non-linear effects by aggregating trees with TMs in the leaves,
and \pkg{\deeptrafo} supports both splines from \pkg{mgcv} and neural networks
from \pkg{tensorflow}. Package \pkg{\deeptrafo} allows for scalable optimization
via mini-batch training.
}\label{tab:tram}
\end{table}

\paragraph{Neural network-based transformation models}
With the advent of (deep) neural networks and the routine collection of
non-tabular data, the idea to combine deep learning and distributional
regression approaches was adopted in several ways. For instance,
\citet{Ruegamer.2020} parameterize distributional regression models via neural
networks, \citet{sick2020deep} describe flexible deep transformation models for
continuous responses. \citet{kook2020ordinal} focus on semi-structured
regression for ordinal responses, and \citet{rugamer2021timeseries} extend the
DCTM approach to distributional autoregressive models for time series responses.
Alternative approaches to combining regression with neural networks including
generalized additive models for location, scale, and shape have been implemented
in \pkg{deepregression} \citep{pkg:deepregression}. In this paper, we present
\pkg{\deeptrafo}, which unifies the above DCTM approaches in a single
\proglang{R} package. 

\paragraph{Comparison to existing packages}
Combining distributional regression with neural network-based estimation has
many advantages, such as modularity (data analysts can easily use
well-established problem-specific neural network architectures), and easy
handling of big datasets (\eg through mini-batch gradient descent with adaptive
learning rates). Thus, like \pkg{tram}, \pkg{\deeptrafo} relies on maximizing a
likelihood function. However, stochastic first-order optimization, such as
stochastic gradient descent and the ability to deal with non-tabular data
distinguishes the two packages (Table~\ref{tab:tram}). Further, \pkg{\deeptrafo}
covers and extends models implemented in \pkg{cotram}. Like \pkg{tramME},
\pkg{\deeptrafo} also allows the specification of smoothing splines via
\pkg{mgcv} \citep{pkg:mgcv}. However, the focus of our package does not lie on
random effects. Penalization as in \pkg{tramnet} is also available for
\pkg{\deeptrafo}. Lastly, unlike models in \pkg{deepregression} do not require
specification of a parametric family of distributions for the response given
covariates.

The rest of this paper is organized as follows. Section~\ref{sec:dctm}
introduces the statistical theory behind TMs and DCTMs. The inner workings of
\pkg{\deeptrafo} are described in Section~\ref{sec:package}, where several case
studies on how to setup up, fit, validate, and interpret DCTMs are presented. We
present an application to binary classification with tabular and text
modalities, and an application to time series modeling via autoregressive TMs
\citep{rugamer2021timeseries}. The appendix contains information on advanced
usage of the package, \eg how censored responses are handled
(Appendix~\ref{app:cens}) or how to warmstart or fix parameters of certain
predictors (Appendix~\ref{app:warm}). In Appendix~\ref{app:fac}, we demonstrate
the package for large tabular datasets and factors with many levels, which
cannot be handled by standard implementations of classical regression models.

\subsection{Deep conditional transformation models}\label{sec:dctm}

\begin{figure}[t!]
\center
\includegraphics[width=0.9\textwidth]{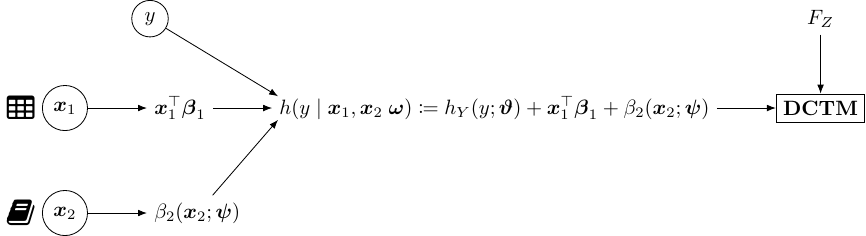}
\includegraphics[width=1\textwidth]{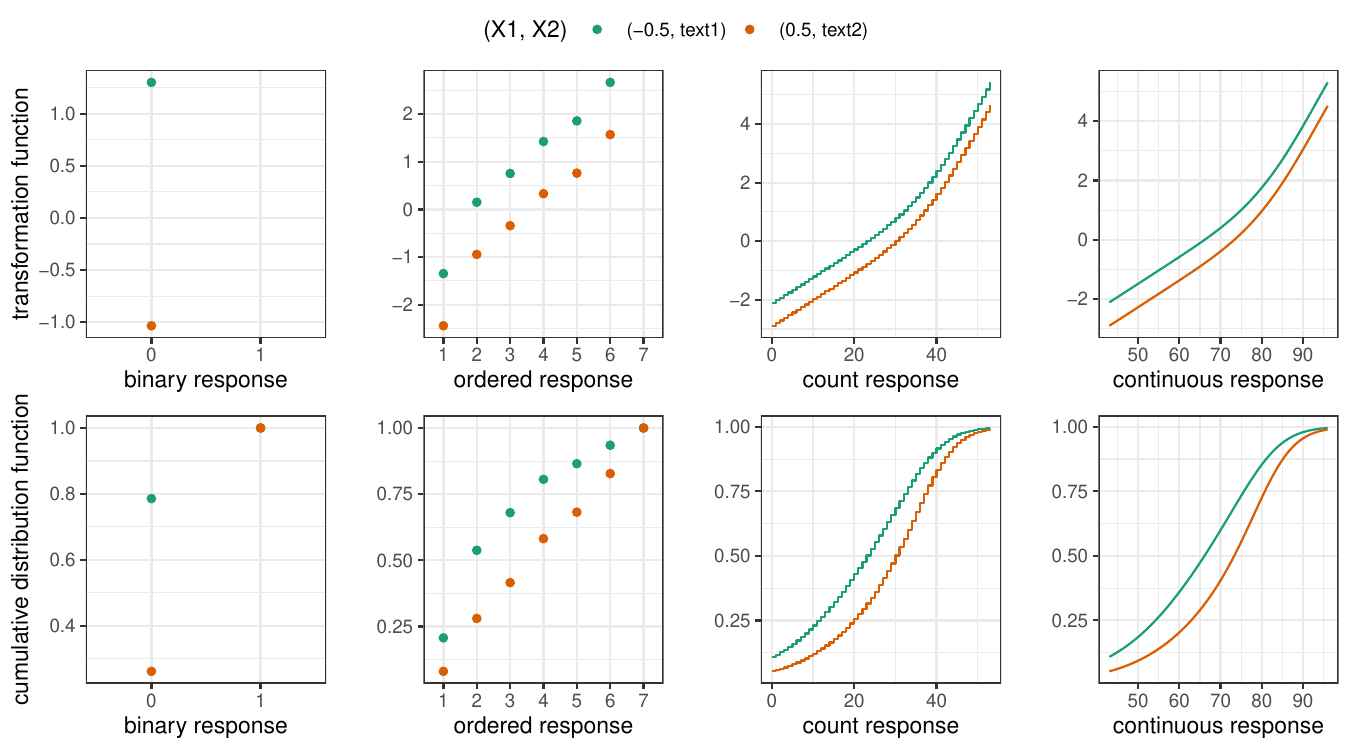}
\caption{Example of a \dctm{} with transformation function $\h(\ry \given \rx_1,
  \rx_2)$ depending on a tabular modality $\rx_1$ and a text modality $\rx_2$,
  which both enter as an additive shift term. The tabular modality enters as a
  simple linear predictor $\rx_1^\top\shiftparm_1$ and the text data via the
  output of a neural network $\beta_2$, which is specified by the user. Together
  with a baseline transformation $\hY$, whose parameterization is discussed
  later, and the latent distribution $\pZ$, the DCTM is fully specified. On the
  bottom, the transformation function $\h$ and cumulative distribution function
  $F_{\rY \given \rX_1 = \rx_1, \rX_2 = \rx_2} = \pZ \circ \h$ are depicted for
  a binary, ordered, count, and continuous response for two realizations of the
  tabular (\textsf{X1}, \textsf{X2}) and text modalities (\textsf{text1},
  \textsf{text2}). For binary and ordered responses with $K$ levels, the
  transformation function contains one and $K - 1$ parameters, respectively,
  because the CDF is constrained to one for the largest class.
}\label{fig:extrafo}
\end{figure}

Transformation models \citep{hothorn2014conditional,hothorn2018most} estimate
the conditional cumulative distribution function (CDF) of a response $\rY \in
\calY \subseteq \RR$ given a realization $\rx$ of covariates $\rX \in\calX$,
\begin{align}
    \pYx(\ry) \coloneqq \Prob(\rY \leq \ry \given \rX = \rx) 
\end{align}
without committing to a particular parametric family of distributions for
$\pYx$. Instead of estimating the CDF directly, transformation models estimate
how to transform the response (conditional on covariates) to a latent variable
$Z \coloneqq \h(\rY\given\rx)$ (independent of $\rX$) with fixed and
user-defined CDF $\pZ : \RR \to [0,1]$, using the \emph{transformation function}
$\h : \calY \times \calX \to \RR$, which is constrained to be monotonically
non-decreasing in $\ry \in \calY$ for all $\rx\in\calX$. Then, the conditional
CDF of the outcome given covariates can be evaluated using the latent CDF $\pZ$
and the transformation function $h$:
\begin{align}
    \Prob(\rY \leq \ry \given \rX = \rx) = \Prob(\h(\rY \given \rx) \leq 
        \h(\ry\given\rx) \given \rX = \rx) = \Prob(Z \leq \h(\ry \given \rx))
        = \pZ(\h(\ry \given \rx)).
\end{align}
For continuous responses, $\h$ is continuous and for discrete responses, $\h$ is
discrete (see Figure~\ref{fig:extrafo}). Expressing the conditional CDF in terms
of $\pZ$ and $\h$ yields simple expressions for probability density and mass
functions and thus also the log-likelihood.

Depending on the choice of $\pZ$ and restrictions on the functional form and
parameterization of $\h$, TMs cover a wide range of well-known models with
varying complexity. 
\begin{example}[Beyond normal linear regression]
Choosing $\pZ = \Phi$ and $\h(\ry \given \rx) = \sigma^{-1}(\ry - \alpha -
  \rx^\top\shiftparm)$, with standard deviation $\sigma > 0$, and intercept
  $\alpha \in \RR$, is equivalent to a normal linear regression model, since
  $\Prob(\rY \leq \ry \given \rX = \rx) = \Phi(\sigma^{-1}(\ry - \alpha -
  \rx^\top\shiftparm))$. Fixing the transformation function to be linear will
  always result in conditionally normal outcome distributions. However, this
  restriction can be lifted by using a non-linear increasing transformation,
  $\hY : \calY \to \RR$, \ie $\h(\ry \given\rx) = \hY(\ry) -
  \rx^\top\widetilde\shiftparm$, which now assumes that the transformed response
  $\hY(\rY)$ (instead of the original response) is normal with mean
  $\rx^\top\widetilde\shiftparm$.
\end{example}

\begin{example}[Beyond Weibull regression]
Choosing $\pZ(\rz) = 1 - \exp(-\exp(\rz))$ with $\h(\ry \given \rx) = a + b
  \log\ry + \rx^\top\shiftparm$, with intercept $a$ and slope $b > 0$, is
  equivalent to a Weibull regression model, since $\Prob(\rY \leq \ry \given \rX
  = \rx) = 1 - \exp(-\exp(a + b \log\ry + \rx^\top\shiftparm) = 1 -
  \exp(-\widetilde{a}\ry^b\exp(\rx^\top\shiftparm))$, where $\widetilde{a}
  \coloneqq \exp(a)$. Also in this example, log-linearity of the transformation
  function fixes the conditional outcome to be Weibull distributed. Allowing an
  arbitrary increasing function, $\hY : \calY \to \RR$, instead, \ie $\h(\ry
  \given\rx) = \hY(\ry) + \rx^\top\widetilde\shiftparm$, results in the Cox
  proportional hazards model, since the survivor function equals $\Prob(\rY \geq
  \ry \given \rX = \rx) = \exp(-\exp(\hY(\ry)\exp(\rx^\top
  \widetilde\shiftparm)))$ and $\exp(\hY(\ry))$ is the cumulative baseline
  hazards.
\end{example}

Thus, TMs contain both normal linear and Weibull regression but also extend both
to a more flexible counterpart that does not assume a parametric family of
conditional outcome distributions.

\paragraph{Parameterizing the transformation function}
In semi-structured regression, we have access to $J$ input modalities $\rx_1,
\dots, \rx_J$, such as tabular data, images, or text, from which we construct
structured (\eg linear, sparse, or smooth) or unstructured (\eg neural network)
predictors. These inputs may be non-tabular, \ie there may be a $j$ for which
$\rx_j \in \calX_j \not\subseteq \RR^d$. By $\calX \coloneqq \calX_1 \times
\dots \times \calX_J$ we denote the entire input space. In \dctm{s},
restrictions on the functional form of $\h$, \ie the way predictors are
constructed based on the input data, lead to varying degrees of interpretability
and flexibility of the model. We begin with an example before introducing $\h$
in its most flexible form. Consider a problem with a single tabular ($\rx_1 \in
\calX_1 \subseteq \RR^p$) and a single text modality ($\rx_2 \in \calX_2$). Data
analysts commonly assume additivity in the effects the separate modalities,
which can be realized by modelling the effect of both modalities as shift terms,
\begin{align}\label{eq:example}
    \h(\ry \given \rx_1, \rx_2; \omegavec) = \hY(\ry; \parm) + \rx_1^\top\shiftparm_1 + 
    \beta_2(\rx_2;\psivec),
    \quad \ry \in \calY,
\end{align}
where $\hY:\calY \to \RR$ denotes the baseline transformation (\ie the
transformation function obtained when $\rx_1^\top\shiftparm_1 + \beta_2(\rx_2) =
0$, which is parameterized in terms of $\parm \in \RR^M$). Further,
$\shiftparm_1$ denotes the coefficients of the linear predictor and $\beta_2 :
\calX_2 \to \RR$ denotes the unstructured predictor for the text data, which are
typically controlled by a neural network with weights $\psivec$. By $\omegavec
\coloneqq (\parm, \shiftparm, \psivec)$, we denote the collection of all
parameters, including the neural network weights. A \dctm{} with
(\ref{eq:example}) is distribution-free because for any constellation of
covariates for which the shifting predictor is zero, \ie for all $(\rx^0_1,
\rx^0_2) \in S^0 \coloneqq \{(\rx_1, \rx_2) \in \calX_1 \times \calX_2 \given
\rx_1^\top\shiftparm_1 + \beta_2(\rx_2) = 0\}$, and all conditional
distributions $Y \given \rX_1 = \rx^0_1, \rX_2 = \rx^0_2$, there exists a unique
baseline transformation given by $\hY = \pZ^{-1} \circ F_{\rY \given \rX_1 =
\rx^0_1, \rX_2 = \rx^0_2}$. In (\ref{eq:example}), covariate effects are assumed
to enter additively on the scale of the transformation function, thus
restricting distributions that can be modeled for $(\rx_1, \rx_2) \in \calX
\backslash S^0$. This argument can be extended to more complex \dctm{s}
\citep[for shift-scale see, \eg][]{siegfried2022distribution}. The example in
(\ref{eq:example}) is depicted in Figure~\ref{fig:extrafo} for typical types of
responses and standard logistic latent distribution.

In \pkg{\deeptrafo}, the most general transformation function $\h$ is
parameterized in terms of $\omegavec \coloneqq (\parm, \shiftparm, \phivec,
\psivec) \in \RR^{Md} \times \RR^p \times \RR^q \times \RR^s$ which serves as
the collection of parameters for basis expansions (potentially including neural 
networks) of the response and input modalities,
\begin{align} \label{eq:trafo}
   \h(\ry \given \rx; \omegavec) = \left(\basisy(\ry) \otimes 
   \basisx(\rx; \phivec) \right)^\top\parm + 
   \svec(\rx; \psivec)^\top\shiftparm, \quad \ry \in \calY,
   \ \rx \in \calX,
\end{align}
where $\otimes$ denotes the Kronecker product and $\basisy : \calY \to \RR^M,
\basisx : \calX \to \RR^d, \svec : \calX \to \RR^p$ denote the bases for the
response, and the $J$ predictors, which either interact ($\basisx(\bcd;
\phivec)$) with the response or simply shift ($\svec(\bcd; \psivec)$) the
transformation function. The dimensions of the neural network weights $\phivec$
and $\psivec$ depend on the complexity of the neural network architectures which
the user has full control over. In \pkg{\deeptrafo}, the basis for the response
is not data-dependent and thus contains no parameters. The interacting and
shifting basis, however, depend on the covariates and may include splines or
neural networks, whose parameters are collected in $\phivec$ and $\psivec$,
respectively.

The transformation function is required to be monotonically non-decreasing for
all $\rx\in\calX$. Hence, depending on the choice of basis, the parameters
$\parm$ in (\ref{eq:trafo}) need to fulfill positivity or monotonicity
constraints \citep{hothorn2014conditional}, which can be enforced by appropriate
reparameterizations. Without interacting predictors, Bernstein polynomials and
discrete bases require $\eparm_1 \leq \eparm_2 \leq \dots \leq \eparm_M$ and
linear and log-linear bases require positive slopes. For more complex
interacting predictors, the positivity of $\basisx(\bcd; \phivec)$ has to be
enforced together with more complex constraints on $\parm$ to ensure a
monotonically non-decreasing transformation function \citep[for details
see][]{baumann2020deep}.

Shift effects are constant across all values of the response, \ie the
transformation $\h$ can only shift up- or downwards (see
Fig.~\ref{fig:extrafo}). The effect of interacting predictors may vary with the
response and thus the shape of  $\h$ may change for different predictor values.
For instance, an interacting binary predictor leads to two separate
transformations for each level, much like stratum variables in survival analysis
allow for separate hazard functions \citep{collett2015modelling}. However, in
its general form, interacting predictors may also include neural networks and
thus unstructured predictors, making them extremely versatile. Scale effects as
introduced in \citet{siegfried2022distribution} are a special case of
interacting predictors, which are included in \pkg{\deeptrafo} by using $\basisx
: \calX \to \RR_+$ with $\rx \mapsto \sqrt{\exp(\gamma(\rx))}$ and, \eg a neural
network $\gamma : \calX \to \RR$. With a linear basis $\basisy$ in $\ry$,
$\basisy(\ry) = (1, \ry)^\top$, this is equivalent to location-scale regression
with error distribution $\pZ$.

\paragraph{Supported response types}
Several types of univariate, potentially censored, responses can be handled.
This includes continuous ($\calY \subseteq \RR$), survival ($\calY \subseteq
\RR_+$), count ($\calY = \mathbb{N}$), and ordered ($\calY = \{\ry_1, \dots,
\ry_K\}$) responses. For continuous responses, the basis for $\rY$ is a smooth
function parameterized via polynomials in Bernstein form of order $M-1$, denoted
by $\bern{M-1}(\ry)$. For count responses ($M=K$), the polynomials in Bernstein
form are evaluated only at the integers, \ie $\bern{M-1}(\lfloor\ry\rfloor)$
\citep{pkg:cotram}. For ordered responses, a dummy-encoding is used, \ie for $k
= 1, \dots, K$, $\bvec(\ry_k) = \evec_k$, where $\evec_k$ denotes the $k$-th
unit vector. Linear and log-linear bases are supported as well. In
Appendix~\ref{app:basis}, we describe how the user can supply custom basis
functions.

\paragraph{Fitting transformation models}
Finally, transformation models can be fitted by minimizing the negative average
log-likelihood over the class of transformation functions $\h(\ry\given\rx;
\omegavec)$ with parameters $\omegavec$,
\begin{align}\label{eq:nll}
    \NLL(\omegavec; \ry_i, \rx_i) \coloneqq - \frac{1}{n} \sum_{i = 1}^n 
    \ell(\omegavec; \ry_i, \rx_{i}),
\end{align}
where the observations $\{(\ry_i, \rx_{i})\}_{i=1}^n$ are assumed to be (conditionally)
independent. In \pkg{\deeptrafo}, the default optimizer is (stochastic) gradient
descent using Adam \citep{Kingma2015adam}. However, any \pkg{keras}
\citep{pkg:keras} or \pkg{tensorflow} optimizer or a custom optimization routine
can be used instead. For a single observation $(\ry, \rx)$, the log-likelihood
contribution $\ell(\h; \ry, \rx)$ depends on the type of censoring of the
observed response. Exact responses $\ry$ contribute
$\log\dZ(\h(\ry\given\rx))\h'(\ry \given \rx)$ to the log-likelihood.
Interval-censored responses $(\ubar\ry, \bar\ry]$ contribute
$\log(\pZ(\h(\bar\ry\given\rx)) - \pZ(\h(\ubar\ry \given \rx)))$. Left- and
right-censored observations follow from the interval-censored contribution as a
special case, by letting $\ubar\ry \to -\infty$ and $\bar\ry \to +\infty$,
respectively \citep{hothorn2014conditional}. In \pkg{deeptrafo}, the
log-likelihood contributions are implemented in terms of mathematical operations
implemented in \pkg{tensorflow}, which call their \proglang{Python} counterpart
via \pkg{reticulate} and allow efficient computation of the log-likelihood, its
gradients and weight updates during optimization.

\subsection{Autoregressive transformation models}\label{sec:atm}

Time series data pose one particular case where the independence assumption
between observations is not tenable and needs to be taken into account.
Formally, the joint distribution of a time series $(Y_t)_{t\in \mathcal{T}}$
with $\mathcal{T} \subseteq \mathbb{N}_0$ can always be factorized in its
conditional distributions, \ie by conditioning $Y_t$ on its full history
$\calF_{t,1} \coloneqq (Y_{t-1}, \ldots, Y_{1})$. A simplification is to impose
a Markov property of order $p \geq 1$ which implies that the conditional
distribution of $Y_t$ only depends on the history up to and including $t-p$,
that is $\calF_{t,t-p} \coloneqq (Y_{t-1}, \ldots, Y_{t-p})$ rather than the
entire history $\calF_{t,1}$. 

Package \pkg{\deeptrafo} offers three ways on how to model time series data
assuming the Markov property. The naive way is given by classical transformation
models where $\calF_{t,t-p}$ is regarded in the basis expansion of the
transformation function shown in \eqref{eq:trafo} where elements of
$\calF_{t,t-p}$ may interact with the response $Y_t$ and simultaneously shift
the transformation function. Furthermore, \cite{rugamer2021timeseries} proposed
the class of autoregressive transformation models (ATMs) which differ from the
naive approach (\ie classical transformation models) in two perspectives. First,
the transformation function $h_t$ in ATMs can be time-varying which may result
in different transformations for different sub-periods. Second, the same $h_t$
is applied to $Y_t$ and each element of $\calF_{t,t-p}$ simultaneously,
resulting in a shared transformation between $Y_t$ and its lags. 

A special subclass of ATMs are AT($p$) models which do not allow for interacting
elements of $\calF_{t,t-p}$ with $Y_t$ through $\basisx$ but restrict to a
linear shift impact of the transformed values of $\calF_{t,t-p}$ on the scale of
$h$. The class of AT($p$) models is closely related to a well-known class of
time series models, \ie autoregressive models of order $p$
\citep[AR($p$),][]{hamilton2020}. In fact, AT($p$) models are equivalent to
AR($p$) models for $P=1$, $\svec(\rx) \equiv \rx$ and the independent white
noise follows the distribution $F_Z$ \citep[for details
see][]{rugamer2021timeseries}. Learning the transformation simultaneously for
the response and its lags as it is done in ATMs is particularly important for
ordinal time series, for which the dimensionality of the model can thereby be
reduced. Instead of modeling each level of the lagged response, the
one-dimensional transformed lagged response is included. It also allows for a
more consistent interpretation in the sense of autoregression because we model
$h(Y_t)$ at the current step (auto)regress the next time point  $h(Y_{t+1})$ on
the likewise transformed response $\h(\rY_t)$, not on the untransformed $Y_t$.
We showcase the practical differences between linear transformation models,
AT($p$) and ATM models in Section~\ref{subsec:ATM}.

\subsection{Application datasets}\label{sec:dataset}

\begin{figure}[t!]
\begin{minipage}[c]{0.59\textwidth}
\begin{tabular}{|p{0.9\textwidth}}
``superman returns discover 5 absence allowed lex luthor walk free closest 
abandoned moved luthor plots ultimate revenge millions killed change planet 
forever ridding steel''
\end{tabular}
\end{minipage}
\begin{minipage}[c]{0.39\textwidth}
\includegraphics[width=1\textwidth]{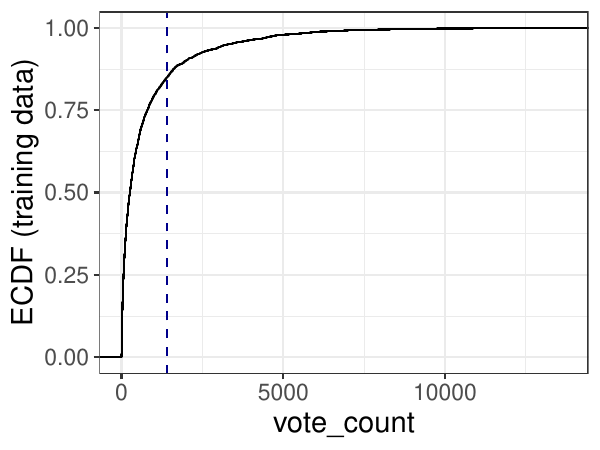}
\end{minipage}
\caption{%
Left: The pre-processed movie review of a picked instance of the 
\code{movies} dataset, in which stop words and punctuation have already 
been removed. Right: The empirical CDF of \code{vote\_count} over all 
movies in the training, where the picked instance has a 
\code{vote\_count} of 1400 as indicated by the dashed line (placing it 
above the top quartile). The used tabular input data comprise 
\code{popularity} (4.08 for the picked instance) and
\code{revenue} (\$400~million for the picked instance).
}
\label{fig:movie}
\end{figure}

\paragraph{Movies data}
In Section~\ref{sec:package}, we will illustrate the features of \pkg{\deeptrafo} 
using the \code{movies} dataset \citep{movies}. The dataset contains information 
on 45,000 movies released prior to July 2017, including number of ratings, budget, 
revenue, popularity, run time, and genre. In addition, non-tabular reviews of the 
movies are available as text data.
In Section~\ref{sec:package}, we will focus on estimating the conditional
distribution of \code{vote\_count} given whether a movie is an action movie,
its budget, its popularity score, and the text review. In Section~\ref{sec:ontram},
we will switch to the binary classification task of deciding whether a movie falls 
into the action genre or not. This way, we can showcase how to apply \dctm{s} for
a wider range of outcome types. 
We pre-process budget, revenue, and popularity using $\log(1 + x)$, due to
their skewed nature. 
In Figure~\ref{fig:movie}, we show the empirical CDF of the variable 
\code{vote\_count} of the \code{movies} 
dataset and provide more information on the used variables for one specific movie.
For the text data, we use a \code{text\_tokenizer} with a
1,000 word vocabulary, convert text to sequence and pad sequences to a maximum
length of 100 and truncate the end of a review.
We use such a simple embedding to illustrate the key steps of the analysis
and make the computations feasible on a standard laptop with 8~gigabytes of
RAM. We additionally present results with a pre-trained embedding that
performs comparably in terms of test NLL in Appendix~\ref{app:word2vec}.

\paragraph{Temperature data} An application of autoregressive transformation
models to a time series of monthly mean maximum temperature in Melbourne (Australia)
in degrees Celsius between January 1971 and December 1990 (240 records) is
presented in Section~\ref{sec:atm}. The \code{temperature} time series was recorded 
by the Australian Bureau of Meteorology and later provided in \citet{Hyndman2022}. 

\section{The package} \label{sec:package}

\begin{figure}[t!]
    \center
    \includegraphics[width=0.85\textwidth]{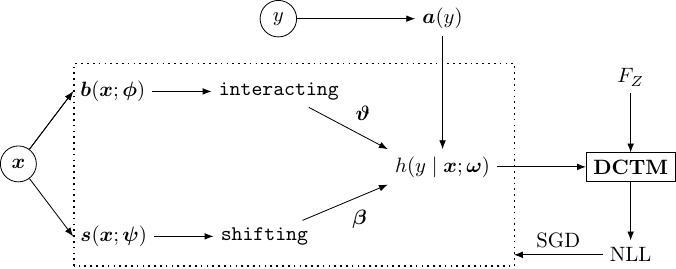}
    \caption{Schematic depiction of setting up and fitting DCTMs. Bases for
    input predictors $\rx$ and response $\ry$ (circles) are evaluated and 
    enter the two neural network components \code{interacting} and 
    \code{shifting} according to \eqref{eq:annotatedtrafo}. The components'
    outputs make up the transformation function $\h$ 
    which is parameterized in terms of $\omegavec$.
    Together with the latent distribution $\pZ$, the loss, \eg NLL, and its
    gradients can be evaluated and used to update parameters $\omegavec$. 
    Since $\pZ$ is parameter-free, all trainable parameters are in the
    transformation function, as indicated by the dotted box.
    }
    \label{fig:deeptrafo}
\end{figure}

Package \pkg{\deeptrafo} builds upon \pkg{tensorflow} as a fitting engine and
\pkg{deepregression} for setting up 
structured model terms such as 
linear effects or splines within a neural network. In contrast to 
\pkg{deepregression}, which implements models with parametric families and 
individual additive predictors, \pkg{\deeptrafo} supports more complicated 
computations such as in \eqref{eq:trafo}. This is exposed to the user via 
\pkg{\deeptrafo}'s formula interface.
In \pkg{\deeptrafo{}}, response, interacting, and shifting terms are represented
as \cls{formula} objects and correspond to the bases in \eqref{eq:trafo}. 
Internally, a \code{processor} is defined for each model term, which evaluates
its basis functions and optional penalties
via \pkg{deeptrafo} internal, \pkg{mgcv}, or \pkg{keras}/\pkg{tensorflow} functions.
For instance, for a continuous response, a polynomial basis in Bernstein form 
and its derivatives are set up by default (\cf Table~\ref{tab:mods}).
The corresponding basis functions are implemented in \pkg{\deeptrafo}.
Package \pkg{\deeptrafo} can include 
terms modeled by user-specified neural network architectures for the 
interacting and shifting terms (see Figure~\ref{fig:extrafo} and 
Figure~\ref{fig:deeptrafo}). 
When initializing the model using such a formula-based call, the model is 
internally translated to \pkg{tensorflow} computations using a computational 
graph.
In the end, a single end-to-end trainable neural network 
is set up, which may contain different neural network components for different 
terms in the interacting or shifting predictor. Together with the supplied 
\code{latent_distr} $\pZ$, the DCTM is fully specified and its parameters can be
estimated by minimizing the NLL via stochastic gradient descent (SGD). 
Since the DCTM has internally been translated to a model from \pkg{tensorflow}, 
the optimization can be done using the \pkg{keras} API, which implements the SGD
routine with many choices for adaptive learning rates while providing training
metrics without requiring users to define training loops for parameter updates.
An appropriate last-layer transformation ensures monotonicity constraints of the 
interacting model term in the response. 

\paragraph{Workflow}
Typical workflows around \pkg{deeptrafo}, including the illustration in
Section~\ref{sec:package} and both applications on binary classification
(Section~\ref{sec:ontram}) and distributional time series
(Section~\ref{subsec:ATM}), are structured as follows: First, a model formula is
set up. The \cls{formula} object encodes in which way each feature enters the
model. If neural network components are used, the corresponding architectures
have to be defined beforehand. Next, the latent distribution $\pZ$ is chosen and
decides which scale the partial effects of components in the formula are
interpreted. Although the formula together with the latent distribution formally
specify the TM completely (Figure~\ref{fig:deeptrafo}), the data and optimizer
have to be supplied at this stage. For deep learning models (as opposed to
statistical models), it is common to separate model building from model fitting,
in order to supply more arguments (such as callbacks) to the optimization
routine. Now, hyperparameters can be tuned based on cross-validation. Finally,
with the chosen hyperparameters, either a single instance of the \dctm{} or an
ensemble is fitted and can be used for downstream prediction tasks. In
Section~\ref{subsec:maincomp}, we describe each step of the workflow in more
detail using the \code{movies} data.

Each step in the \pkg{\deeptrafo} workflow is highly customizable, \eg custom
functions for basis evaluation (Appendix~\ref{app:basis}), custom last-layer
transformations, and general-purpose optimization routines
(Section~\ref{sec:ontram}), such as SGD with adaptive learning rates
(Appendix~\ref{app:opt}), can be supplied. 

\begin{table}[!ht]
    \centering
    \resizebox{0.98\textwidth}{!}{%
    \begin{tabular}{lrrr}
    \toprule
        \bf Model function & \bf Model name & \bf Default basis & \bf Default latent distribution \\
    \midrule
        \code{BoxCoxNN} & Transformed normal & Bernstein & Standard normal \\
        \code{ColrNN} & Continuous outcome logistic & Bernstein & Standard logistic \\
        \code{cotramNN} & Count transformation & Bernstein & Standard logistic \\
        \code{CoxphNN} & Cox proportional hazards & Bernstein & Standard minimum extreme value \\
        \code{LehmannNN} & Lehmann-type & Bernstein & Standard maximum extreme value \\
        \code{LmNN} & Normal linear & Linear & Standard normal \\
        \code{PolrNN} & Proportional odds logistic & Discrete & Standard logistic \\
        \code{SurvregNN} & Weibull & Log-linear & Standard minimum extreme value
    \\ \bottomrule
    \end{tabular}}
    \caption{Supported models together with the default choice of basis function
    and latent distribution. Model functions summarized here are implemented 
    with a specific choice of basis and latent distribution that define
    commonly applied regression models.} \label{tab:mods}
\end{table}

\subsection{Main components}\label{subsec:maincomp}

We describe the main components of \pkg{\deeptrafo} below by showing how to use 
the formula interface, set up a DCTM, and fit the model. In this section, all 
steps are illustrated with the \code{movies} dataset. 
In the following examples, we assign non-default values to some of the arguments 
that can be supplied to
functions and methods for building and fitting \pkg{keras}-based neural networks. 
This is not because the models have been tuned extensively, but rather to illustrate
the most important hyperparameters that are involved in building and fitting
DCTMs.

\subsubsection{Formula interface} 

Models can be specified via a formula interface akin to the one used in 
\pkg{tram} \citep{pkg:tram}, where covariates interacting with the response 
are supplied on the left-hand side, and shift effects are supplied on 
the right-hand side of the formula, as illustrated below.
\begin{CodeChunk}
\begin{CodeInput}
response | interacting ~ shifting
\end{CodeInput}
\end{CodeChunk}
Thus, the formula interface mimics the transformation function as introduced in
\eqref{eq:trafo}:
\begin{align}\label{eq:annotatedtrafo}
   \big(
   \underbrace{\basisy(\ry)}_{\mathtt{response}}
   \otimes 
   \overbrace{\basisx(\rx; \phivec)}^{\mathtt{interacting}}
   \big)^\top\parm + 
   {\underbrace{\svec(\rx;\psivec)}_{\mathtt{shifting}}}^\top\shiftparm.
\end{align}

\textbf{Case study: Formula interface}

We begin by modeling the conditional distribution of \code{vote\_count}
given a binary indicator of whether the movie is categorized as an action
movie or not (\code{genreAction}), the movie's \code{budget} and its
\code{popularity}. The below formula allows for separate baseline 
transformations of the response for action movies \emph{vs.}~all other
genres, a smooth effect for \code{budget} and a linear effect for 
\code{popularity}. Here, we use the standard spline basis representation
implemented in \pkg{mgcv}. A non-exhaustive list of smoothers and other 
processors is given in Table~\ref{tab:covariates}. Processors are
specialized functions for handling predictors which can speed up 
computation. For instance, \code{fac_processor()} from \pkg{safareg}
evaluates factors on-line and thus facilitates computation for large
factor models \citep[][also see the illustration in 
Appendix~\ref{app:fac}]{ruegamer2022fastr}.
\begin{CodeChunk}
\begin{CodeInput}
R> fm <- vote_count | genreAction ~ 0 + s(budget, df = 3) + popularity
\end{CodeInput}
\end{CodeChunk}
In the above formula we exclude an additional intercept in the shift 
term by specifying \mbox{\code{0 + ...}}, because the interacting basis 
already contains an intercept.

\begin{table}[!ht]
    \centering
    \begin{tabular}{ll}
    \toprule
        \bf Effect / Processor &  \bf Example formula \\ \midrule
        Linear &  \code{y} \msim \code{x} \\
        Smooth &  \code{y} \msim \code{s(x, ...)} \\
        Tensor product splines &  \code{y} \msim \code{[te|ti|t2](x, ...)} \\
        Lasso &  \code{y} \msim \code{lasso(x)} \\
        Group lasso &  \code{y} \msim \code{grlasso(x)} \\ 
        Row-wise tensor product &  \code{y} \msim \code{rwt(x)} \\ 
        \midrule
        Neural network &  \code{y} \msim \code{nn(x)} \\ 
        \midrule
        Processor &  \code{*_processor} \\
                  &  \eg \code{fac_processor} 
     \\ \bottomrule
    \end{tabular}
    \caption{Implemented choices of \code{interacting} and \code{shift} 
    processors. All splines from \pkg{mgcv} are supported. Custom neural networks 
    can be supplied as functions via \lodm{}. Additional processors, for example, 
    for faster processing of large factors or interactions from \pkg{safareg}, can 
    be included via \code{additional\_processors} 
    \protect{\citep{ruegamer2022fastr,ruegamer2022afm}}. All terms can also be
    included as interacting effects on the left-hand side of the formula, \eg
    \code{y | term(x, ...)} \msim \code{1}.}\label{tab:covariates}
\end{table}

\subsubsection{Setting up DCTMs}

DCTMs can be generically set up using the \code{deeptrafo()} function.
\begin{CodeChunk}
\begin{CodeInput}
deeptrafo(formula = response | interacting ~ shifting, data = ...)
\end{CodeInput}
\end{CodeChunk}
The \code{data} can be supplied as a \code{data.frame} or \code{list}. The
function returns a \cls{deeptrafo} object, whose methods are described in
Section~\ref{subsec:methods}.

Special cases of DCTMs coincide with well-known models and are given their
own function in \pkg{\deeptrafo}. The naming conventions in \pkg{\deeptrafo}
follow the \pkg{tram} package \citep{pkg:tram} and add the suffix \code{NN}.
For instance, the proportional odds logistic regression model (ordinal response
and $\pZ = \expit$) is implemented as \code{Polr()} in \pkg{tram} and \code{PolrNN()}
in \pkg{\deeptrafo} (see Table~\ref{tab:mods} for an overview). 

\textbf{Case study: Setting up DCTMs}

For the \code{movies} data, we set up a count transformation model with standard
logistic latent distribution. The logistic distribution is chosen, so that the
partial effects of the features are interpretable as log-odds ratios. Example
interpretations are given in Section~\ref{sec:ontram}. We supply the Adam
optimizer (the default, see Appendix~\ref{app:opt}) for SGD with learning rate
of 0.1 decaying with a rate of $4 \cdot 10^{-4}$ \citep{Kingma2015adam}. The
training data \code{train} is the result of the preprocessing steps described in
Section~\ref{sec:dataset}. The code for reproducing all output and figures can
be found on \proglang{GitHub} at
\url{https://github.com/LucasKook/case-study-deeptrafo.git}.
\begin{CodeChunk}
\begin{CodeInput}
R> opt <- optimizer_adam(learning_rate = 0.1, decay = 4e-4)
R> (m_fm <- cotramNN(formula = fm, data = train, optimizer = opt))
\end{CodeInput}
\begin{CodeOutput}
	 Untrained count outcome deep conditional transformation model

Call:
cotramNN(formula = fm, data = train, optimizer = opt)

Interacting:  vote_count | genreAction 

Shifting:  ~0 + s(budget, df = 6) + popularity 

Shift coefficients:
s(budget, df = 6)1 s(budget, df = 6)2 s(budget, df = 6)3 s(budget, df = 6)4 
             0.557             -0.702              0.760             -0.181 
s(budget, df = 6)5 s(budget, df = 6)6 s(budget, df = 6)7 s(budget, df = 6)8 
            -0.201             -0.687              0.670              0.671 
s(budget, df = 6)9         popularity 
            -0.377             -0.888 
\end{CodeOutput}
\end{CodeChunk}
Printing the model to the console shows the model specification and shift 
coefficients. Note that the model has only been randomly initialized and 
not yet fitted, as indicated by ``Untrained count outcome deep conditional 
transformation model'' in the \code{print()} call. Upon calling \code{fit()},
\code{ensemble()}, or \code{cv()}, the model's history will be non-empty
and it will be considered ``trained'' when printed again.

\subsubsection{Fitting DCTMs}

For fitting DCTMs the user calls \code{fit()}, which calls the model
internal \code{mod\$fit_fun()}, per default a wrapper around 
\code{fit.keras.engine.training.Model()}, with the supplied arguments
(for instance \code{epochs}, \code{batch\_size}). All functionalities
of fitting \code{keras} models carry over to fitting \dctm{s}, including
callbacks (\ie custom operations applied after every iteration or mini-batch
update).

\textbf{Case study: Fitting DCTMs}

\begin{figure}[t!]
    \centering
    \includegraphics[width=0.85\textwidth]{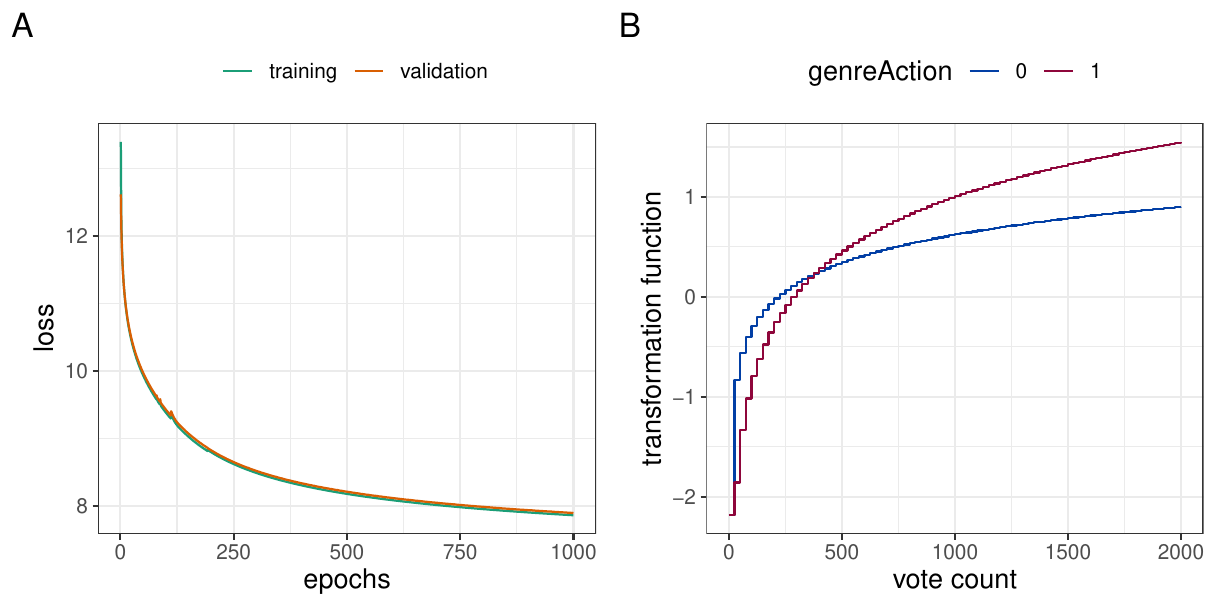}
    \caption{%
    A: Training and validation loss trajectory for \code{m\_fm}.
    B: Estimated transformation functions for both levels of
    \code{genreAction} with \code{popularity} and 
    \code{budget} fixed at their mean in the training
    data.
    }
    \label{fig:mfmhist}
\end{figure}

The \cls{\deeptrafo{}} object returned by \code{cotramNN} is fitted for 1,000
epochs, with a batch size of 64, and a 10\% validation split. The validation
split is used during training to judge whether overfitting occurs
\citep{goodfellow2016deep}. Below, we print the (now trained) model.
\begin{CodeChunk}
\begin{CodeInput}
R> m_fm_hist <- fit(m_fm, epochs = 1e3, validation_split = 0.1, 
+    batch_size = 64, verbose = FALSE)
R> unlist(coef(m_fm, which = "shifting"))
\end{CodeInput}
\begin{CodeOutput}
s(budget, df = 6)1 s(budget, df = 6)2 s(budget, df = 6)3 s(budget, df = 6)4 
           0.38339           -0.28824           -0.04608           -0.03992 
s(budget, df = 6)5 s(budget, df = 6)6 s(budget, df = 6)7 s(budget, df = 6)8 
           0.00616           -0.02692           -0.00511            0.01355 
s(budget, df = 6)9         popularity 
          -0.36587           -0.82771 
\end{CodeOutput}
\end{CodeChunk}
Figure~\ref{fig:mfmhist}A depicts the training and validation loss trajectory
for inspecting convergence and overfitting, which can be generated with
\code{plot(m\_fm\_hist)}.
The learning curves indicate that the model is not fully trained after 1000
epochs and there is no evidence for overfitting.
Figure~\ref{fig:mfmhist}B shows the estimated transformation function. 
In Section~\ref{subsec:methods}, we describe how to produce
plots of the transformation function and density. Since \code{genreAction} is
included as a response-varying effect, the two transformation functions are 
allowed to cross.

\subsubsection{Working with neural networks}

The \pkg{\deeptrafo} package allows to directly model effects of, for instance, 
text or image data via neural networks.
In \dctm{s}, neural networks map from a complex input space, such as text or
images, to Euclidean space. When the neural network enters as a shift term,
the output of the network is a real number which is interpretable on the 
latent scale $\pZ^{-1}$, \ie the scale of the transformation function. Custom
neural networks can be supplied to \code{deeptrafo} as functions or 
\cls{keras\_model}s via the \lodm{} argument.

\textbf{Case study: Working with neural networks}

In our running example, we use the following architecture to model the contribution of 
the movie reviews provided as textual descriptions. In Section~\ref{sec:ontram}, we 
present an application with further downstream analysis of the text embedding
and how this simple embedding compares against using larger pre-trained
embeddings.
\begin{CodeChunk}
\begin{CodeInput}
R> embd_mod <- function(x) x |>
+    layer_embedding(input_dim = nr_words, output_dim = embedding_size) |>
+    layer_lstm(units = 50, return_sequences = TRUE) |>
+    layer_lstm(units = 50, return_sequences = FALSE) |>
+    layer_dropout(rate = 0.1) |> layer_dense(25) |>
+    layer_dropout(rate = 0.2) |> layer_dense(5) |>
+    layer_dropout(rate = 0.3) |> layer_dense(1)
\end{CodeInput}
\end{CodeChunk}
The neural network \code{embd_mod} maps movie ratings to a real value (for more
details see the case study in Section~\ref{sec:ontram}). The interpretational 
scale of output depends on the choice of latent distribution. Here, the
logistic distribution ($\pZ = \expit$) renders the output of \code{embd\_mod}
interpretable on the log-odds scale. In turn, differences in the output of
\code{embd\_mod} can be interpreted as log odds-ratios when changing, for
instance, a single word in a sentence and leaving everything else constant. 
In our \deeptrafo{} model, we can now supply a named list 
\code{list(deep = embd\_mod)} and use \code{deep(texts)} in the formula.
\begin{CodeChunk}
\begin{CodeInput}
R> fm_deep <- update(fm, . ~ . + deep(texts))
R> m_deep <- cotramNN(fm_deep, data = train, 
+    list_of_deep_models = list(deep = embd_mod))
R> fit(m_deep, epochs = 50, validation_split = 0.1, batch_size = 32,
+    callbacks = list(callback_early_stopping(patience = 5)), 
+    verbose = FALSE)
\end{CodeInput}
\end{CodeChunk}
The default optimization routine may not produce optimization paths as 
smooth as when omitting the neural network component. However, adaptively scheduled learning 
rates for SGD often work well out-of-the-box, \eg using \code{optimizer = optimizer\_adam()} 
as an argument when initializing the \cls{\deeptrafo{}} model. Sometimes also different 
learning schedules are needed for the different modalities (see Section~\ref{sec:ontram}).

\subsubsection{Ensembling DCTMs}

A simple and popular method to improve prediction performance and to quantify
training stability (\ie uncertainty from random initialization and stochastic
optimization) are deep ensembles \citep{lakshminarayanan2016deepensembles}.
In a deep ensemble, a neural network model is trained $B$ times using the same training 
and validation data, but different initial weights. Training via SGD may then converge
to different (local) minima and the members may yield different predictions. However,
averaging the predicted densities of the $B$ ensemble members is guaranteed to improve
upon the average individual performance (\eg in terms of NLL). In \pkg{\deeptrafo},
an ensemble of a model can be fitted via \code{ensemble()}. Besides classical deep
ensembling, \pkg{\deeptrafo} implements transformation ensembles \citep{kook2022deep}.
Transformation ensembles are specifically tailored towards DCTMs and preserve their 
additive structure and thus (partial) interpretability by averaging the predicted 
transformation functions instead of the predicted densities.

\textbf{Case study: Ensembling \dctm{s}}

\begin{figure}[!t]
    \centering
    \includegraphics[width=0.6\textwidth]{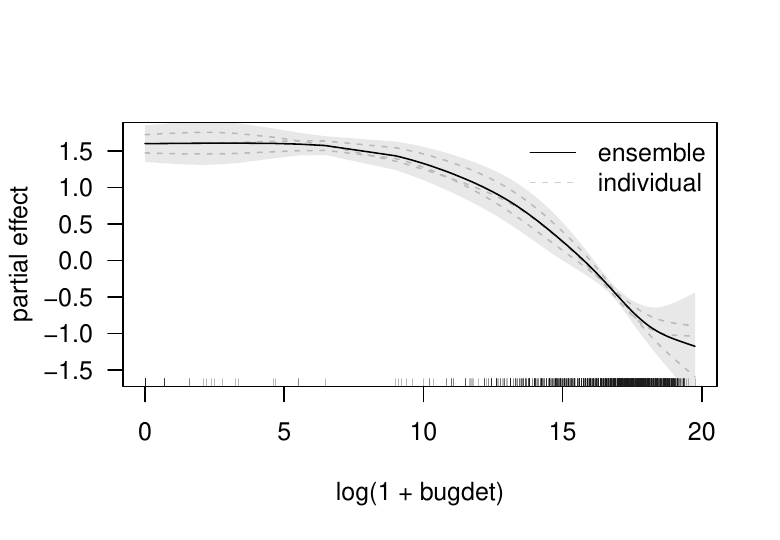}
    \caption{Training stability in the estimated smooth partial effect 
        of \code{budget} on \code{vote\_count} obtained via transformation 
        ensembling.}
    \label{fig:ens}
\end{figure}

Below, we fit five instances of \code{m\_deep}. Then, we combine their 
predictions on the scale of the transformation function and can investigate
uncertainty in the effects of the shifting predictors and prediction performance
on the test set.
\begin{CodeChunk}
\begin{CodeInput}
R> ens_deep <- ensemble(m_deep, n_ensemble = 3, epochs = 50, batch_size = 64,
+    verbose = FALSE)
\end{CodeInput}
\end{CodeChunk}
Figure~\ref{fig:ens} shows the estimated smooth effect of \code{budget}
with training stability indicated by the shaded area. Investigating the
out-of-sample prediction performance, we see that the transformation
ensemble performs better than the members do on average
\citep[see Proposition~3 in][]{kook2022deep}.
\begin{CodeChunk}
\begin{CodeInput}
R> unlist(logLik(ens_deep, convert_fun = \(x) - mean(x)))
\end{CodeInput}
\begin{CodeOutput}
members1 members2 members3     mean ensemble
   -8.28    -8.50    -8.32    -8.37    -8.35
\end{CodeOutput}
\end{CodeChunk}

\subsubsection{Cross-validating DCTMs for hyperparameter tuning}

With \code{cv()}, \pkg{\deeptrafo} provides a cross-validation function for 
\cls{\deeptrafo} objects. When supplying an integer to \code{cv_folds}, the
data is split into \code{cv_folds} number of folds. Alternatively, the user
can specify a list with two elements indicating data indices for training
and validation. The output of \code{cv()} can be used for tuning 
smoothing hyperparameters, choosing between including a predictor as 
interacting or shifting, or different neural network architectures.

\textbf{Case study: Cross-validating \dctm{s}}

\begin{figure}[!t]
    \centering
    \includegraphics[width=0.9\textwidth,%
        trim={0.1cm 0.2cm 1cm 1cm},clip]{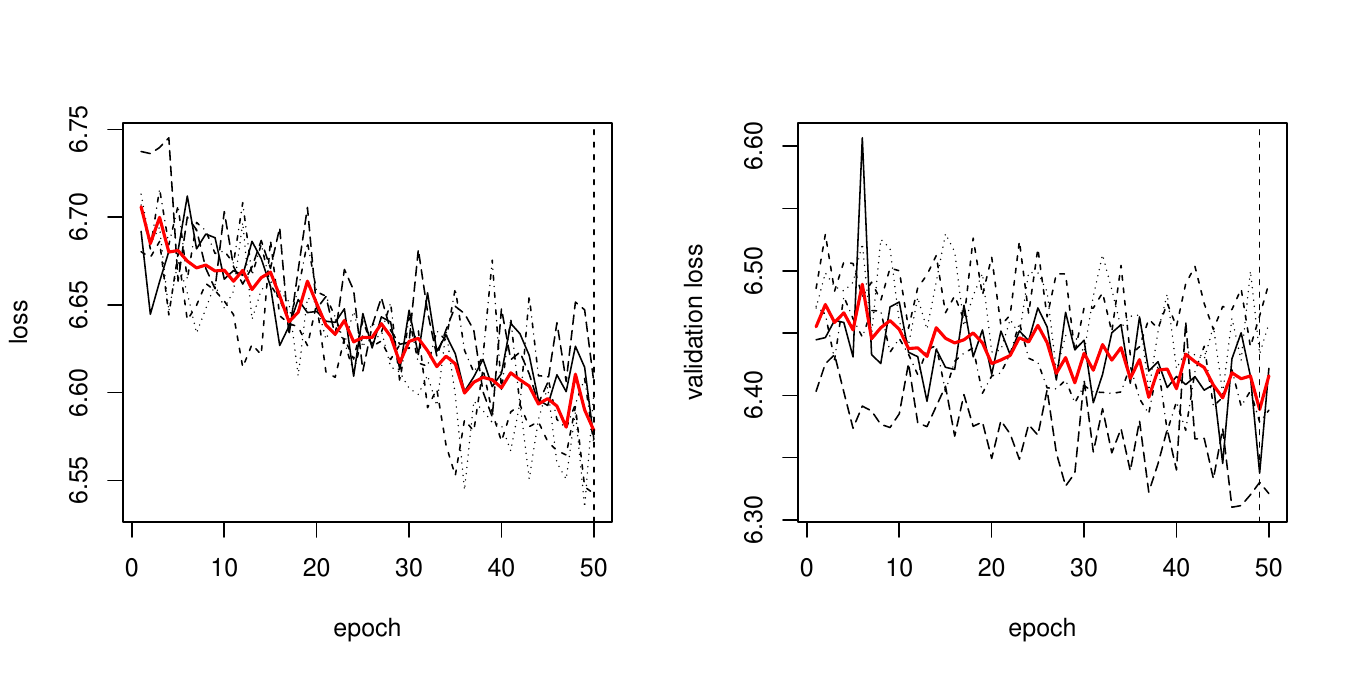}
    \caption{Default plot generated by \code{cv.deeptrafo()}. The vertical
    lines indicate the epoch with minimal average train/validation loss.}
    \label{fig:cv}
\end{figure}

The following call to \code{cv()} performs 5-fold cross validation while 
fitting each instance of \code{m_deep} for 50 epochs. Train and validation
loss trajectories are shown in Figure~\ref{fig:cv}. The vertical bars
indicate the epoch with the best average train/validation loss.
\begin{CodeChunk}
\begin{CodeInput}
R> cv_deep <- cv(m_deep, epochs = 50, cv_folds = 5, batch_size = 64) 
R> plot_cv(cv_deep)
\end{CodeInput}
\end{CodeChunk}

\subsection{Methods overview} 
\label{subsec:methods}

In the following, we briefly describe \proglang{S}3~methods for \cls{deeptrafo}
and \cls{dtEnsemble} objects.

\subsubsection[Methods for `deeptrafo' objects]{Methods for \cls{deeptrafo} objects}

\begin{itemize}
    \item \code{coef} return coefficients for the interacting
        or shifting terms (controllable via
        \linebreak
        \mbox{\code{which\_param = c("shifting", "interacting", "autoregressive")}}).
    \item \code{predict} returns in-sample predictions when
        \code{newdata} is not supplied. The supported types are \code{"trafo"},
        \code{"pdf"}, \code{"cdf"}, \code{"interaction"}, \code{"shift"},
        \code{"terms"}. When \code{newdata} is supplied, predictions are
        evaluated at the response, if it is contained in \code{newdata}.
        The response can be omitted from \code{newdata} to predict the whole
        conditional distribution. Then, predictions are evaluated on a grid 
        of length \code{K}, which is automatically generated based on the 
        response's support in the training data set. A custom grid of response 
        values can be supplied via \code{q}, which overwrites \code{K}.
    \item \code{logLik} evaluates in- or out-of-sample 
        log-likelihoods. This can be useful for model criticism and
        evaluating predictive performance, respectively. The argument
        \code{convert\_fun} controls how the individual NLL contributions
        are summarized. The default is 
        \linebreak
        \mbox{\code{function(x) = -sum(x)}} to 
        compute the log-likelihood. Other common choices include \code{identity}
        to obtain the individual NLL contributions, or \code{mean} to get the 
        average NLL.
    \item \code{plot} by default plots smooth components in
        the \code{shifting} formula part. Data for plotting can be obtained
        by setting \code{only\_data = TRUE}. Smooth terms in 
        \code{interacting} can be plotted by setting 
        \code{which\_param = "interacting"}. In the same manner as in
        \code{predict}, densities evaluated in-sample (\code{type = "pdf"}),
        CDFs (or probability integral transforms, with \code{type = "cdf"}),
        and transformation functions (\code{type = "trafo"}) can be obtained.
        When omitting the response from \code{newdata}, the whole density,
        cumulative distribution, or transformation function can be plotted.
    \item \code{print} prints a brief summary of the \dctm{}
        including coefficients of additive linear and smooth terms in
        \code{shifting}. Setting \code{with\_baseline = TRUE} also prints
        coefficients of linear and smooth terms in \code{interacting}.
        The \code{print\_model} argument toggles whether the \pkg{keras}
        summary of the \dctm{} should be printed in addition.
\end{itemize}

\subsubsection[Methods for ``dtEnsemble'' objects]{Methods for \cls{dtEnsemble} objects}

Methods \code{coef} and \code{predict} of \cls{deeptrafo} objects take the same
arguments as their 
\linebreak
\cls{deeptrafo} counterparts. The output is returned for
all ensemble members. Likewise, \code{logLik} returns the processed NLL
contributions for individual ensemble members, their average, and the
transformation ensemble.

\section{Application: Binary classification}\label{sec:ontram}

In this application, we use the \code{movies} dataset and fit four different 
models with the goal to predict the binary response \code{action} (0:~non-action 
movie, 1:~action movie, defined in the next code chunk),
which encodes whether a movie is an action movie or not. The model \code{m_0} 
is unconditional; \code{m_tab} uses only one tabular predictor, 
\code{popularity}, as linear shift predictor; \code{m_text} uses only 
\code{texts} as an unstructured shift predictor; \code{m_semi} is a 
semi-structured model which uses both modalities as shift predictors.
The purpose of the analysis is to show the potential gains in prediction
performance that can be achieved when including the text data and learning
an embedding, for which conventional statistical models would require
extensive feature engineering. The models that do not include the text
data could, in principle, also be fitted using \code{glm()} from the
\pkg{MASS} \citep{pkg:MASS} package and yield virtually the same results 
as \pkg{\deeptrafo}.

First, we encode the binary response as an ordered factor allowing us to use 
the framework of ordinal neural network transformation models 
\citep{kook2020ordinal}. 
This step is necessary because unordered factors are not supported by
\pkg{\deeptrafo}.
\begin{CodeChunk}
\begin{CodeInput}
R> train$action <- ordered(train$genreAction)
R> test$action <- ordered(test$genreAction, levels = levels(train$action))
\end{CodeInput}
\end{CodeChunk}

We then set up the formulas for the four models. The unconditional model is
specified without any predictors and \code{1} on the right-hand side. Later,
we will restrict this additional intercept to zero for identification (see
\code{warmstart_weights} in the definition of \code{m_0}). For all other
models, we remove the intercept directly by specifying \code{0 + ...} on
the right-hand side.
\begin{CodeChunk}
\begin{CodeInput}
R> fm_0 <- action ~ 1
R> fm_tab <- action ~ 0 + popularity
R> fm_text <- action ~ 0 + deep(texts)
R> fm_semi <- action ~ 0 + popularity + deep(texts)
\end{CodeInput}
\end{CodeChunk}
Here, \code{deep} is the same neural network architecture with text 
embedding as in Section~\ref{subsec:maincomp}. We use a custom
\cls{keras\_model} to which we can refer to by the name \code{"embd"}
and wrap it in a function, which allows us to create multiple instances
of the same model. This may be necessary in applications to make the
resulting \proglang{Python} objects point to different copies in memory
and avoid unintented re-use of already trained or modified models.
\begin{CodeChunk}
\begin{CodeInput}
R> make_keras_model <- function() {
+    return(keras_model_sequential(name = "embd") |> 
+    layer_embedding(input_dim = nr_words, output_dim = embedding_size) |>
+    layer_lstm(units = 50, return_sequences = TRUE) |> 
+    layer_lstm(units = 50, return_sequences = FALSE) |> 
+    layer_dropout(rate = 0.1) |> layer_dense(25) |>  
+    layer_dropout(rate = 0.2) |>  layer_dense(5, name = "penultimate") |>
+    layer_dropout(rate = 0.3) |> layer_dense(1))
+  }
\end{CodeInput}
\end{CodeChunk}

Next, we use \code{PolrNN()} to set up the different models with a standard 
logistic latent distribution. Models including text data are trained
for ten epochs with early stopping and a patience of two, and the 
weights from the epoch with the best validation loss are restored. The
unconditional and tabular-only models are trained full-batch and without
validation split until converging to the minimum 
since convexity of the problem implies a unique solution.

Besides the simple text embedding that is trained from scratch, we also
present how to use a pre-trained \code{word2vec} embedding in 
Appendix~\ref{app:word2vec}.

\subsection{Unconditional model}

For the unconditional model, the intercept is fixed to zero via 
\code{warmstart_weights} to ensure identification. The details 
explaining the next code chunk can be found in Appendix~\ref{app:warm}.
\begin{CodeChunk}
\begin{CodeInput}
R> m_0 <- PolrNN(fm_0, data = train, optimizer = optimizer_adam(
+    learning_rate = 1e-2, decay = 1e-4), weight_options = weight_control(
+    general_weight_options = list(trainable = FALSE, use_bias = FALSE),
+    warmstart_weights = list(list(), list(), list("1" = 0))))
R> fit(m_0, epochs = 3e3, validation_split = 0, batch_size = length(
+    train$action), verbose = FALSE)
\end{CodeInput}
\end{CodeChunk}

The unconditional model \code{m_0} has one parameter which estimates the log-odds 
of a movie belonging to a non-action genre without any predictors. The estimated intercept
parameter, given by \code{coef(m_0, which = "interacting")}, corresponds to the single 
(fix) value of the transformation function $\h$ for a binary response 
(see Figure~\ref{fig:extrafo}). The code chunk below shows that
the estimated intercept agrees with the observed log-odds of a movie belonging
to a non-action genre up to numerical inaccuracies.
\begin{CodeChunk}
\begin{CodeInput}
R> all.equal(unlist(unname(coef(m_0, which = "interacting"))), 
+    qlogis(mean(train$action == 0)), tol = 1e-6)
\end{CodeInput}
\begin{CodeOutput}
[1] TRUE
\end{CodeOutput}
\end{CodeChunk}

We can obtain the unconditional log-odds also using \code{predict()} with 
\code{type = "trafo"}. From the estimated log-odds we can determine the 
probability for a movie to belong to a non-action genre which matches 
the prevalence of non-action movies in the train set. The prevalence of non-action 
movies can also be computed directly by using the \code{predict} function and setting 
the argument \code{type = "pdf"} and supplying \code{action = 0} in \code{newdata}.

\subsection{Tabular-only model}

Next, we set up and fit \code{m_tab} including \code{popularity} as a linear shift
predictor.
\begin{CodeChunk}
\begin{CodeInput}
R> m_tab <- PolrNN(fm_tab, data = train, optimizer = optimizer_adam(
+    learning_rate = 0.1, decay = 1e-4))
R> fit(m_tab, epochs = 1e3, batch_size = length(train$action),
+    validation_split = 0, verbose = FALSE)
\end{CodeInput}
\end{CodeChunk}

We obtain the estimated linear shift parameter $\hat\beta$ of \code{m_tab} by 
\code{coef(m_tab, which_param = "shifting")}. Here, the odds for
a movie to belong to genre action change by the factor $\exp(-\hat\beta)$
\begin{CodeChunk}
\begin{CodeInput}
R> exp(-unlist(coef(m_tab, which = "shifting")))
\end{CodeInput}
\begin{CodeOutput}
popularity
      1.54
\end{CodeOutput}
\end{CodeChunk}
when the predictor \code{popularity} increases by one unit. Without flipping the
sign, the coefficient $\hat\beta$ represents a log-odds ratio for a movie
belonging to a non-action genre compared to genre action upon a one-unit change
in \code{popularity}. Thus, the interpretation of $\hat\beta$ depends on the
parameterization of the model, in particular, the sign of the shifting predictor.
In \pkg{\deeptrafo{}}, the shifting predictor is consistently parameterized with
a plus sign for all models, which may differ from other implementations of the
same model type (\eg generalized linear models in \pkg{MASS}, or TMs in \pkg{tram}).

\subsection{Text-only model}

We now define and fit \code{m_text} including only the tokenized movie reviews. 
\begin{CodeChunk}
\begin{CodeInput}
R> embd <- make_keras_model()
R> m_text <- PolrNN(fm_text, data = train, list_of_deep_models = list(
+    deep = embd), optimizer = optimizer_adam(learning_rate = 1e-4))
R> fit(m_text, epochs = 10, callbacks = list(callback_early_stopping(
+    patience = 2, restore_best_weights = TRUE)), verbose = FALSE)
\end{CodeInput}
\end{CodeChunk}
Analogously to smooth partial effects, the differences between two shift estimates 
resulting from two different text inputs can still be interpreted as log odds-ratios.

We now have a closer look at what the \code{embd_mod} has learned. The network
takes as input the words (encoded as indices). Here, we use a vocabulary (all 
words in the data set) of 10000 words and limit each review text to a size of 
100 words. Review texts which are shorter are prepended with zeros, longer 
movie descriptions are cut after 100 words. All punctuation is removed.
The \code{layer_embedding} learns to embed the word indices into an 
\code{embedding_size}-dimensional representation. The resulting word embeddings 
of the text are the input sequence to an LSTM layer with a 50-dimensional 
memory state. The second LSTM layer outputs the 50-dimensional state after the last 
word in the text, which is then further processed by a fully connected neural 
network including dropout to prevent overfitting. 

We can now use the trained \code{embd} to extract and analyze the derived 
latent features of the embedding of single words or whole texts. We can obtain
the embedding of a single word as the output of \code{layer_embedding()}.
If we use a whole review as input, the latent features in the layer 
\code{"penultimate"} correspond to a five-dimensional representation of
the text embedding processed by subsequent layers. 

\begin{figure}[!t]
    \centering
    \includegraphics[width=\textwidth]{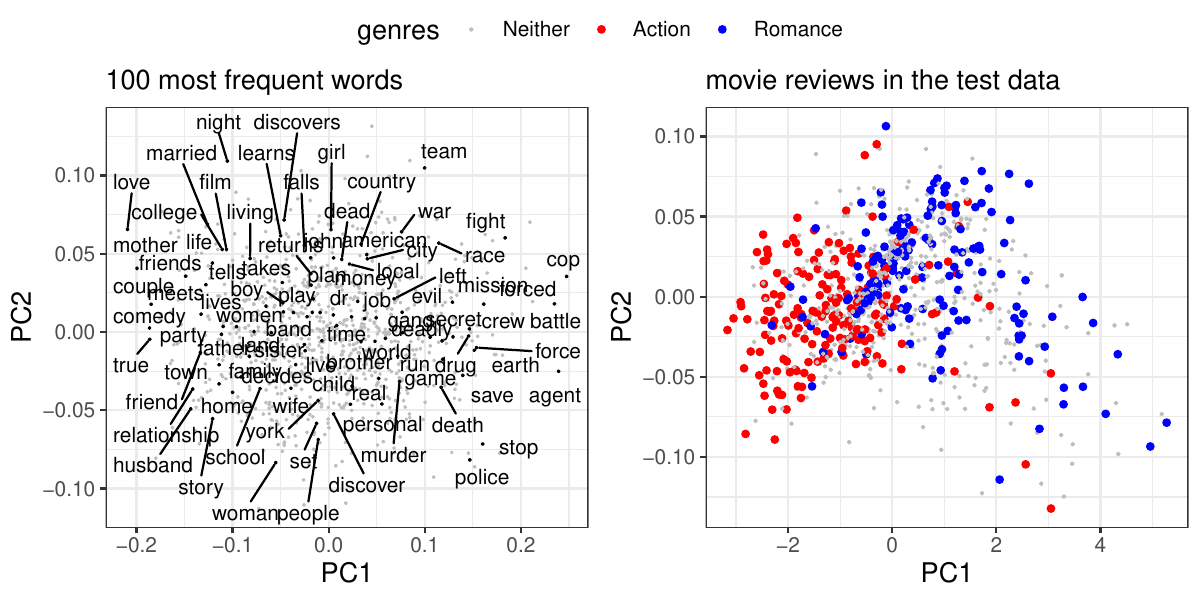}
    \caption{%
    The first two principal components of the \code{"embedding"} layer for 
    single words (left) and the lower-dimensional representation learned in 
    \code{"penultimate"} (right) for whole movie reviews in the test data.
    On the left, the PCA is computed on the word embedding of the 1,000 most 
    frequent words (we display only the 100 most frequent, black dots), and
    on the right based on the low-dimensional representation of the embedding 
    of the 888 full movie reviews contained in the test data.
    }\label{fig:figure_movie_pca}
\end{figure}

Figure~\ref{fig:figure_movie_pca} shows the first two components of a 
principle component analysis (PCA) applied to the word embedding (left) and
to the features learned in the \code{penultimate} layer for whole reviews
(right). The left plot reveals, that words hinting at an action movie, have a
similar embedding, and are separated from words that are rather representative
of a romance movie. The plot on the right of Figure~\ref{fig:figure_movie_pca}
confirms that the features derived from the embedding are tailored to discriminate 
action movies from other genres since latent features of action movies cluster 
together and are fairly well separated from romantic movies.

\subsection{Semi-structured model} \label{subsec:semi}

Finally, we set up the most complex model \code{m_semi} which takes both data 
modalities as input. To achieve efficient training of the tabular part and 
avoid overfitting of the embedding network \code{emdb_semi} we use two different 
learning rates for the structured and unstructured part of the model. Specifically,
we optimize the intercept (with name \code{"ia_1__2"}) and tabular shift predictor
(with name \code{"popularity_3"}) with a higher learning rate, than the embedding 
model (\code{"embd"}). In the embedding model, some layers are named explicitly,
the names for the other components can be obtained from the \cls{keras\_model}
summary by initializing and calling \code{print(m_semi, print\_model = TRUE)}.

\begin{CodeChunk}
\begin{CodeInput}
R> embd_semi <- make_keras_model()
R> optimizer <- function(model) {
+    optimizers_and_layers <- list(
+      tuple(optimizer_adam(learning_rate = 1e-2),
+      get_layer(model, "ia_1__2")),
+      tuple(optimizer_adam(learning_rate = 1e-2),
+      get_layer(model, "popularity_3")), 
+      tuple(optimizer_adam(learning_rate = 1e-4), 
+      get_layer(model, "embd")))
+    multioptimizer(optimizers_and_layers)
+  }
R> m_semi <- PolrNN(fm_semi, data = train, list_of_deep_models = list(
+    deep = embd_semi), optimizer = optimizer)
R> fit(m_semi, epochs = 10, callbacks = list(callback_early_stopping(
+    patience = 2, restore_best_weights = TRUE)), verbose = FALSE)
\end{CodeInput}
\end{CodeChunk}

\subsection{Model comparison}\label{sec:modcomp}

Comparing the prediction performance of the models (measured in terms of
NLL) indicates that mainly the text modality contains information for separating
action movies from other genres. However, for a
more reliable assessment of this statement, the training schedule should be
optimized further. We compute 95\% bootstrap confidence intervals as a simple
uncertainty measure for the test NLL. 
In Appendix~\ref{app:word2vec}, we illustrate how to use pre-trained embeddings
with a shallow and deeper neural network architecture and obtain comparable
results in terms of out-of-sample NLL. Using pre-trained embeddings may reduce 
computation times and yield comparable predictions, especially when the training 
sample size is small \citep{goodfellow2016deep}.
\begin{CodeChunk}
\begin{CodeInput}
R> bci <- function(mod) {
+    lli <- logLik(mod, newdata = test, convert_fun = identity)
+    bt <- boot(lli, statistic = \(x, d) mean(x[d]), R = 1e4)
+    btci <- boot.ci(bt, conf = 0.95, type = "perc")$percent[1, 4:5]
+    c("nll" = mean(lli), "lwr" = btci[1], "upr" = btci[2])
+ }

R> mods <- list("unconditional" = m_0, "tabular only" = m_tab, 
+   "text only" = m_text, "semi-structured" = m_semi)
R> do.call("cbind", lapply(mods, bci))
\end{CodeInput}
\begin{CodeOutput}
    unconditional tabular only text only semi-structured
nll         0.531        0.516     0.437           0.423
lwr         0.501        0.486     0.390           0.372
upr         0.562        0.549     0.486           0.478
\end{CodeOutput}
\end{CodeChunk}
Like \code{m_tab} the model \code{m_semi} estimates a linear shift parameter for 
\code{popularity} which can also be interpreted as a (conditional) log 
odds-ratio. The parameter goes in the same direction as in \code{m_tab} but has 
a reduced absolute value and is now intepretable as a conditional log-odds
ratio because the text information that is now additionally accounted for.
\begin{CodeChunk}
\begin{CodeInput}
R> c("tabular only" = unlist(unname(coef(m_tab))), 
+    "semi-structured" = unlist(unname(coef(m_semi))))
\end{CodeInput}
\begin{CodeOutput}
   tabular only semi-structured 
          -0.43           -0.32 
\end{CodeOutput}
\end{CodeChunk}
The presented case study is meant to showcase some functionality of the package 
\pkg{\deeptrafo} for binary responses. A \code{PolrNN} model for an ordinal
response that has $K$ levels and yields $K-1$ values for a discrete transformation
function (see Figure~\ref{fig:extrafo}) can be interpreted analogously, \eg linear
shift terms are still interpreted as log odds-ratios
\citep[for details and more examples see][]{kook2020ordinal}. 

\section{Application: Autoregressive transformation models}\label{subsec:ATM}

We now return to ATMs, first discussed in Section~\ref{sec:atm}. One special form 
of ATMs are AT($p$) models. AT($p$) models assume a linear impact of the transformed 
values of $\calF_{t,t-p}$ on the scale of $h$. Because the transformation
is the same as for the response, AT($p$) models thus learn a joint transformation 
of the response and its lags. For an illustration of transformation models 
applied to time series data, the \code{temperature} dataset is used. We aim to estimate 
the conditional distribution of the monthly mean maximum temperature in degrees 
Celsius (°C) in Melbourne (Australia) between January 1971 and December 1990. A 
descriptive analysis of the time series shows a strong seasonal pattern. This 
motivates the application of a flexible approach that allows modeling the quickly 
changing moments of the conditional distribution over time.

In the following, we compare three different forms of autoregressive transformation
models. The most flexible model (ATM) includes the lags as interacting predictors
and transformed lags in the shift term. The AT(3) model only includes the transformed
lags in the shift term. Lastly, the naive \code{ColrNN} model (Colr) includes the lags
as an additive linear term. In addition, every model contains a shift effect for \code{month}.
The ATM and AT(3) model can currently only be fitted using \pkg{\deeptrafo}, whereas
the other two models could be fitted using conventional TMs implemented in \pkg{tram}.
We compare the three models based on their estimated transformation functions and conditional 
densities. We start by creating a factor variable \code{month} for the calendar month as well 
as the lags $Y_{t-p}$, $p = 1, 2, 3$ denoted by \code{y_lag_<p>} for including raw additive
lags. AT($p$) lags are included using the internal \code{atplag()} processor.
\begin{CodeChunk}
\begin{CodeInput}
R> lags <- c(paste0("y_lag_", 1:p, collapse = "+"))
\end{CodeInput}
\end{CodeChunk}
The formula for the ATM model is given as follows. We include all three lags as
interacting predictors on the left-hand side of the formula and specify the \code{atplag}s
on the right-hand side.
\begin{CodeChunk}
\begin{CodeInput}
R> (fm_atm <- as.formula(paste0("y |", lags, "~ 0 + month + atplag(1:p)")))
\end{CodeInput}
\begin{CodeOutput}
y | y_lag_1 + y_lag_2 + y_lag_3 ~ 0 + month + atplag(1:p)
\end{CodeOutput}
\end{CodeChunk}
ATP lags can be conveniently included in the formula by specifying the lags
inside \code{atplag()}. For the AT(3) model, we include the transformed lags 
in the shift but not in the interacting term.
\begin{CodeChunk}
\begin{CodeInput}
R> (fm_atp <- y ~ 0 + month + atplag(1:p))
\end{CodeInput}
\begin{CodeOutput}
y ~ 0 + month + atplag(1:p)
\end{CodeOutput}
\end{CodeChunk}
The third model (Colr) we compare is a \code{ColrNN} model which includes 
the raw lags in an additive shift term.
\begin{CodeChunk}
\begin{CodeInput}
R> (fm_colr <- as.formula(paste0("y ~ 0 + month + ", lags)))
\end{CodeInput}
\begin{CodeOutput}
y ~ 0 + month + y_lag_1 + y_lag_2 + y_lag_3
\end{CodeOutput}
\end{CodeChunk}
After preprocessing, the \code{temperature} dataset is saved in \code{d_ts}. We fix 
the support of the response to \code{min_supp = 10} and \code{max_supp = 30} and 
specify Bernstein polynomials of order \code{P = 6}. We use \code{ColrNN()} 
to specify all models. ATM and AT(3) include \code{atplag}s and the third model, Colr, 
does not.
\begin{CodeChunk}
\begin{CodeInput}
R> mod_fun <- function(fm, d) ColrNN(fm, data = d, 
+    trafo_options = trafo_control(order_bsp = P,
+    support = c(min_supp, max_supp)), tf_seed = 1,
+    optimizer = optimizer_adam(learning_rate = 0.01))
R> mods <- lapply(list(fm_atm, fm_atp, fm_colr), mod_fun)
\end{CodeInput}
\end{CodeChunk}
After defining the models, we proceed with training all three models.
In addition, we include callbacks to reduce the learning rate upon
encountering a plateau in the training loss, to ensure convergence of
the optimization procedure.
\begin{CodeChunk}
\begin{CodeInput}
R> fit_fun <- function(m) m |> fit(epochs = ep, callbacks = list(
+   callback_early_stopping(patience = 20, monitor = "val_loss"),
+   callback_reduce_lr_on_plateau(patience = 5)), batch_size = nrow(d_ts_lag),
+   verbose = FALSE)
R> lapply(mods, fit_fun)
\end{CodeInput}
\end{CodeChunk}

\begin{figure}[!t]
    \centering
    \includegraphics[width=\textwidth]{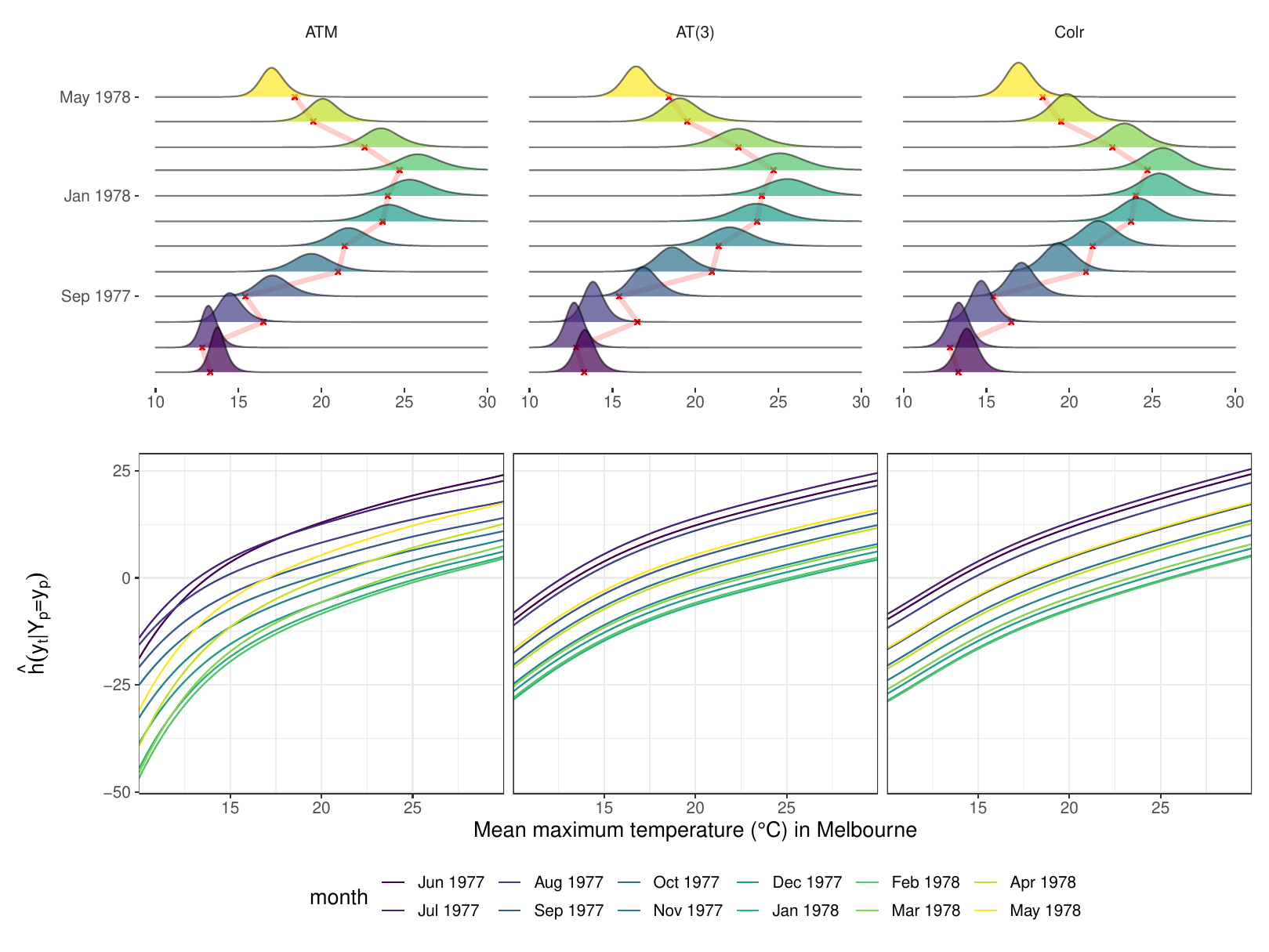}
    \caption{Estimated conditional densities (top row) of monthly temperature
    records between June 1983 and May 1984, based on the ATM model (left), the 
    AT(3) model (center) and the Colr model (right). The observed 
    values across this time span are depicted in red. The plots in the bottom 
    row show the corresponding estimated conditional transformation functions.
    }\label{fig:figure_ATM_trafo}
\end{figure}

We compare the in-sample log-likelihood for the three models for the subset of
data between June 1977 and May 1978 in \code{t_idx}.
\begin{CodeChunk}
\begin{CodeInput}
R> t_span_one <- seq(as.Date("1977-03-01"), as.Date("1978-05-01"),
+    by = "month")
R> ndl <- d_ts[d_ts$time %in% t_span_one]
R> t_span_two <- seq(as.Date("1977-06-01"), as.Date("1978-05-01"),
+    by = "month")
R> ndl_lag <- d_ts_lag[d_ts_lag$time %in% t_span_two]
R> structure(unlist(c(lapply(mods[1:2], logLik, newdata = ndl), 
+    lapply(mods[3], logLik, newdata = ndl_lag))), names = 
+    c("ATM", paste0("AT(", p, ")"), "Colr"))
\end{CodeInput}
\begin{CodeOutput}
  ATM AT(3)  Colr 
-19.5 -22.5 -20.1 
\end{CodeOutput}
\end{CodeChunk}
The comparison shows that the Colr and the ATM model fit similarly well 
compared to the slightly less favorable fit of the AT(3) model. A visual 
inspection of the estimated conditional densities depicted in 
Figure~\ref{fig:figure_ATM_trafo} shows similar results for all three
estimation methods. In summary, the ATM class may be favored over naive 
TMs (Colr) in the time series domain for its autoregressive structural 
assumption, \ie lags entering in a transformed way, identical to the 
transformation of $y_t$ \citep[see][]{rugamer2021timeseries}. 

\section{Conclusion} \label{sec:conclusion}

With \pkg{\deeptrafo}, we introduce the first \proglang{R}~package for fitting a
broad class of distributional regression models with a neural network back-end.
Package \pkg{\deeptrafo} combines the advantages of transformation models, \ie 
flexible distribution-free, yet interpretable models for conditional distributions, 
with the advantages of neural network-based machine learning, which scales well for
large or non-tabular datasets. The intuitive formula interface allows users
familiar with packages such as \pkg{stats} \citep{pkg:stats}, \pkg{MASS},
\pkg{tram}, \pkg{survival} \citep{pkg:survival}, \pkg{mgcv}, and others to easily 
adapt their workflow to neural networks and more complex datasets out-of-the-box.

Users can supply custom basis functions, loss functions, optimization routines 
and neural network architectures to adapt and extend functionalities from 
\pkg{deeptrafo} to problems in which the goal is learning a conditional
cumulative distribution function.
We illustrate \pkg{\deeptrafo} with tabular and text, as well as time series
data with count, discrete, and continuous outcomes, which are all handled in
a unified way. We demonstrate how custom neural network architectures and
optimizers can be used, and how to tune, evaluate, and interpret \dctm{s}.

Applying neural network-based models to analyze text or image data typically
comes with higher flexibility but also larger computational costs compared
to more conventional statistical models. We demonstrate how pre-trained text
embeddings can be used to obtain competitive results to training an embedding
from scatch and reduce the computational and the environmental burden.

\section*{Acknowledgments}
We thank Sandra Siegfried for her valuable comments on the manuscript. The research
of LK was supported by the Swiss National Science Foundation (Grant No.~214457).
LK conducted part of this work at the University of Copenhagen, University of
Zurich and Zurich University of Applied Sciences.
The research of DR has been partially supported by the German Federal Ministry of 
Education and Research (BMBF) under Grant No.~01IS18036A. The research of LK and BS
was supported by Novartis Research Foundation (FreeNovation~2019) and by the Swiss
National Science Foundation (Grant No.~S-86013-01-01 and~S-42344-04-01). The research
of OD has been partially supported by BMBF under Grant No.~01IS19083A. The authors
of this work take full responsibilities for its content.

\bibliography{main}

\appendix

\section*{Appendix}

In the appendix, we describe how \pkg{\deeptrafo} handles censored responses
(Appendix~\ref{app:cens}), how the user can warmstart and fix weights of
interacting and shifting terms (Appendix~\ref{app:warm}), and how to include
custom basis functions (Appendix~\ref{app:basis}).
We demonstrate how to use 
pre-trained embeddings (Appendix~\ref{app:word2vec}) and give details on
the most commonly used options for optimization (Appendix~\ref{app:opt}).
In addition, we describe an 
alternative formula interface (Appendix~\ref{app:form}) and show how to use 
\pkg{\deeptrafo} for large tabular datasets (Appendix~\ref{app:fac}).

\section{Handling censored responses}\label{app:cens}

Package \pkg{\deeptrafo} detects the type of response automatically. However, the user
may specify the type explicitly via \code{response\_type} in \code{deeptrafo()}
and all alias/wrapper functions. Allowed types of responses are continuous,
count, survival, ordered (including binary). Censored responses can be supplied
as \cls{Surv} objects. Internally, ordered and count responses are treated
as censored. For instance, the two observations \code{c(0L, 1L)} with 
\code{response\_type = "count"} are internally represented as left- and 
interval-censored, respectively.

\begin{CodeChunk}
\begin{CodeInput}
R> deeptrafo:::response(y = c(0L, 1L))
\end{CodeInput}
\begin{CodeOutput}
     cleft exact cright cinterval
[1,]     1     0      0         0
[2,]     0     0      0         1
attr(,"type")
[1] "count"
\end{CodeOutput}
\end{CodeChunk}

\section{Warmstarting and fixing weights}\label{app:warm}

Warmstarting and fixing weights may be important in numerical experiments, for
finetuning parts of the models, or transfer learning \citep{goodfellow2016deep}.
In \pkg{\deeptrafo}, the user can supply a \cls{keras\_model}, as returned,
for instance, by \linebreak\code{keras_model_sequential()}. When defining the 
model, \pkg{keras} specific arguments for controlling weight initialization
can be used, as shown below.
\begin{CodeChunk}
\begin{CodeInput}
R> nn <- keras_model_sequential() |>
+   layer_dense(input_shape = 1L, units = 3L, activation = "relu", 
+   use_bias = FALSE, kernel_initializer = initializer_constant(
+   value = 1))
R> unlist(get_weights(nn))
\end{CodeInput}
\begin{CodeOutput}
[1] 1 1 1
\end{CodeOutput}
\end{CodeChunk}

To warmstart or fix coefficients of the interacting or shifting part of
a \dctm{}, the \code{weight\_options} argument in \code{deeptrafo()} can 
supplied with the output of \code{weight\_control()}, which, in addition
to others, takes the same arguments as the \pkg{keras} layers above.
\begin{CodeChunk}
\begin{CodeInput}
R> args(weight_control)
\end{CodeInput}
\begin{CodeOutput}
function (specific_weight_options = NULL, general_weight_options = list(
    activation = NULL, use_bias = FALSE, trainable = TRUE, 
    kernel_initializer = "glorot_uniform", bias_initializer = "zeros",
    kernel_regularizer = NULL, bias_regularizer = NULL, 
    activity_regularizer = NULL, kernel_constraint = NULL,
    bias_constraint = NULL), warmstart_weights = NULL,
    shared_layers = NULL) 
NULL
\end{CodeOutput}
\end{CodeChunk}

Below, we warmstart the shift coefficient for a \code{PolrNN} model.
Here, \code{warmstart\_weights} takes a list with three components,
of which the first two control the weights of the interacting predictor
and the last the weights of the shift predictor. The weights can be
referred to by the name of the covariate, \ie \code{"temp" = 0}.
\begin{CodeChunk}
\begin{CodeInput}
R> data("wine", package = "ordinal")
R> mw <- deeptrafo(
+   response ~ 0 + temp, 
+   data = wine, weight_options = weight_control(warmstart_weights = list(
+     list(), list(), list("temp" = 0))))
R> unlist(coef(mw))
\end{CodeInput}
\begin{CodeOutput}
$temp 
         [,1]
tempwarm    0 
\end{CodeOutput}
\end{CodeChunk}
The three lists correspond to the three formula components
\code{response}, \code{interacting}, and \code{shifting}.
The list corresponding to the response is always 
empty, since it does not contain any parameters. In case there 
is no interacting predictor, the second list corresponds to
the parameters of the basis function of the response, \ie
the intercept function. In case there is no shift term, an
intercept is set up which can be referred to as \code{"1"}
and frozen as illustrated in the main text (Section~\ref{sec:ontram}).
In the example above, we warmstart weights of a component 
in the shift term and supply two empty lists for the other 
components. 

\section{Including custom basis functions}\label{app:basis}

Linear, log-linear, and Bernstein bases, as used by \pkg{\deeptrafo},
require (linear) inequality constraints on their parameters. Internally,
these constraints are handled in \code{trafo\_control()}, by supplying 
an \pkg{keras} layer, which transforms the weights for the interacting
predictor appropriately. In \pkg{\deeptrafo}, the implemented bases
are \code{"bernstein"}, \code{"ordered"}, and \code{"shiftscale"}. The
former two require $\eparm_{jP + 1)} \leq \eparm_{jP + 2} \leq \dots 
\leq \eparm_{jP + P}$, $l = 0, \dots, L - 1$ for $\basisx(\rx) \in 
\RR^L$ and degree $P - 1$ Bernstein basis or ordered response with $P + 1$
levels. The shift-scale basis requires only $\eparm_1 > 0$ in $\ry \mapsto
\eparm_0 + \eparm_1\ry$.

The user can now supply custom basis functions as shown below. First,
the basis (\code{linear_basis}) and its derivative 
(\code{linear_basis_prime}) are defined. Afterwards, the constraints
on the parameters are defined using \proglang{Python}- and 
\pkg{tensorflow}-specified constructs (\code{tf$...}).

\begin{CodeChunk}
\begin{CodeInput}
R> linear_basis <- function(y) {
+    ret <- cbind(1, y)
+    if (NROW(ret) == 1)
+      return(as.vector(ret))
+    ret
+  }
R> linear_basis_prime <- function(y) {
+    ret <- cbind(0, rep(1, length(y)))
+    if (NROW(ret) == 1)
+      return(as.vector(ret))
+    ret
+  }
R> constraint <- function(w, bsp_dim) {
+    w_res <- tf$reshape(w, shape = list(bsp_dim, as.integer(nrow(w) /
+    bsp_dim)))
+    w1 <- tf$slice(w_res, c(0L, 0L), size = c(1L, ncol(w_res)))
+    wrest <- tf$math$softplus(tf$slice(w_res, c(1L, 0L), size = c(
+    as.integer(nrow(w_res) - 1), ncol(w_res))))
+    w_w_cons <- k_concatenate(list(w1, wrest), axis = 1L)
+    tf$reshape(w_w_cons, shape = list(nrow(w), 1L))
+  }
R> tfc <- trafo_control(
+    order_bsp = 1L,
+    y_basis_fun = linear_basis,
+    y_basis_fun_prime = linear_basis_prime,
+    basis = constraint
+  )
\end{CodeInput}
\end{CodeChunk}

We can now compare our re-implementation of a transformation model with
linear basis against \code{Lm()} from \pkg{tram}. To efficiently fit
\dctm{s} for small tabular datasets, we recommend full-batch (\ie batch size
$n$) training with a large learning rate (0.01) and either decay or callbacks
for reducing the learning on validation loss plateaus.
\begin{CodeChunk}
\begin{CodeInput}
R> library("tram")
R> set.seed(1)
R> n <- 1e3
R> d <- data.frame(y = 1 + rnorm(n), x = rnorm(n))
R> m <- deeptrafo(y ~ 0 + x, data = d, trafo_options = tfc,
+    optimizer = optimizer_adam(learning_rate = 1e-2),
+    latent_distr = "normal")
R> fit(m, batch_size = n, epochs = 5e3, validation_split = NULL,
+    callbacks = list(callback_reduce_lr_on_plateau(monitor = "loss")),
+    verbose = FALSE)
R> abs(unlist(coef(m)) - coef(Lm(y ~ x, data = d)))
\end{CodeInput}
\begin{CodeOutput}
       x
0.00017
\end{CodeOutput}
\end{CodeChunk}

\section{Application: Binary classification with pre-trained embeddings}\label{app:word2vec}

As large pre-trained language models become more and more practice in natural 
language processing, we show an alternative way to fit a \dctm{} using a 
pre-trained embedding called \code{word2vec} \citep{mikolov2013efficient}.
The embedding is provided by 
Google and can be downloaded from their servers. Due to its corpus size, the 
embedding file is multiple Gigabytes large. After storing the embedding in the 
\code{./Data/} folder, we can load the embedding using the \pkg{gensim} 
\proglang{Python} library \citep{rehurek2011gensim}
and transform every word in the training dataset into 
a vector in the embedding space.

\begin{CodeChunk}
\begin{CodeInput}
R> embedding_dim <- 300
R> if (file.exists("word2vec_embd_matrix.RDS")) {
R>  embedding_matrix <- readRDS("word2vec_embd_matrix.RDS")
R>  vocab_size <- nrow(embedding_matrix)
R> } else {
R>  gensim <- import("gensim")
R>  model <- gensim$models$KeyedVectors$load_word2vec_format(
+    "../Data/GoogleNews-vectors-negative300.bin", binary = TRUE)
R>  vocab_size <- length(words$word)
R>  embedding_matrix <- matrix(0, nrow = vocab_size, ncol = embedding_dim)
R>  names_model <- names(model$key_to_index)
R>  for (i in 1:vocab_size) {
R>    word <- words$word[i]
R>    if (word %in% names_model) {
R>      embedding_matrix[i, ] <- model[[word]]
R>    }
R>  }
R> saveRDS(embedding_matrix, file = "word2vec_embd_matrix.RDS")
R> }
\end{CodeInput}

Having transformed the text data into vectors in the embedding space, we can
use these in an embedding layer to define our model. We start with a shallow 
neural network that flattens the vectors for each word and learns a linear 
model for the resulting data matrix.

\begin{CodeInput}
R> w2v_mod <- function(x) x |>
+    layer_embedding(input_dim = vocab_size, output_dim = embedding_dim,
+    weights = list(embedding_matrix), trainable = FALSE) |>
+    layer_flatten() |>
+    layer_dense(units = 1)
R> fm_w2v <- action ~ 0 + shallow(texts) 
R> m_w2v <- deeptrafo(fm_w2v, data = train,
+    list_of_deep_models = list(shallow = w2v_mod),
+    optimizer = optimizer_adam(learning_rate = 1e-5))
R> dhist <- fit(m_w2v, epochs = 200, validation_split = 0.1,
+    batch_size = 32, callbacks = list(
+    callback_early_stopping(patience = 5)), verbose = FALSE)
R> bci(m_w2v)
\end{CodeInput}
\begin{CodeOutput}
  nll   lwr   upr 
0.523 0.494 0.553 
\end{CodeOutput}

While stopping the training later may result in further model 
improvement, the negative log-likelihood values already indicate
similar performance to the tabular-only model in Section~\ref{sec:modcomp},
which yielded a test NLL of 0.52. We can improve this model by learning 
a deep neural network on top of the pre-trained embedding using 1D 
convolutions as follows.

\begin{CodeInput}
R> w2v2_mod <- function(x) x |>
+    layer_embedding(input_dim = vocab_size, output_dim = embedding_dim,
+    weights = list(embedding_matrix), trainable = FALSE) |>
+    layer_conv_1d(filters = 128, kernel_size = 5, activation = 'relu') |>
+    layer_max_pooling_1d(pool_size = 5) |>
+    layer_conv_1d(filters = 128, kernel_size = 5, activation = 'relu') |>
+    layer_global_max_pooling_1d() |>
+    layer_dense(units = 128, activation = 'relu') |>
+    layer_dropout(rate = 0.5) |>
+    layer_dense(units = 1)
R> fm_w2v2 <- action ~ 0 + deep(texts)
R> m_w2v2 <- deeptrafo(fm_w2v2, data = train,
+    list_of_deep_models = list(deep = w2v2_mod),
+    optimizer = optimizer_adam(learning_rate = 1e-5))
R> dhist <- fit(m_w2v2, epochs = 200, validation_split = 0.1, 
+    batch_size = 32, callbacks = list(
+    callback_early_stopping(patience = 5)), verbose = FALSE)
R> bci(m_w2v2) 
\end{CodeInput}
\begin{CodeOutput}
  nll   lwr   upr 
0.510 0.486 0.534 
\end{CodeOutput}
\end{CodeChunk}

The test NLL resulting from the deeper architecture is lower
than the one obtained from the shallow architecture above, but
not as low as the one obtained using the text-only model in
Section~\ref{sec:modcomp} of 0.44 (95\% bootstrap confidence
interval from 0.390 to 0.486).

\section{Options for optimization}\label{app:opt}

In deep learning, selecting an appropriate optimizer is crucial for model
performance. If \pkg{\deeptrafo} specifies a model with a deep predictor, exact
optimization is not possible anymore and routines that require second- or
higher-order derivates of the objective are too expensive. Optimization is
therefore done using first-order methods, in particular variations of stochastic
gradient descent (SGD). 

\begin{itemize}
    \item Adam \citep{Kingma2015adam} is widely used due to its effectiveness
      across various applications, offering adaptive learning rates that handle
      sparse gradients efficiently. It is by far the most common choice and
      hence our default option. While Adam's default learning rate and momentum
      parameters can be changed, this must be done with care. 
    \item Another option is SGD with momentum, preferred for optimizing large
      CNNs, with its momentum term accelerating gradients for faster
      convergence. In contrast to Adam, SGD with momentum does not come with a
      well-working default and hence often requires hyperparameter tuning for
      the momentum.
    \item Other notable options include RMSprop \citep{tieleman2012lecture},
      designed for non-stationary objectives and noisy gradients, and Nadam
      \citep{dozat2016incorporating}, which combines elements of Adam and
      Nesterov accelerated gradient. These optimizers are, however, typically
      chosen for specific applications and should be used only after careful
      hyperparameter tuning.
\end{itemize}

Irrespective of the choice of the optimizer, semi-structured models such as DCTM
typically have an imbalance in their optimization dynamic when including deep
neural networks due to the large difference in the number of parameters for
structured and unstructured model components. This can, in particular, lead to
slow convergence of the structured model part. To mitigate this problem, users
can use warm-starts as described in Appendix~\ref{app:warm}, or use optimizers
with different learning rates for the different model components as described in
Section~\ref{subsec:semi}.

\section{Alternative formula interface}\label{app:form}

Following {\sc ontram}s (ordinal neural network transformation models),
introduced in Kook \& Herzog \emph{et al.} (\citeyear{kook2020ordinal}),
\pkg{\deeptrafo} offers an alternative formula interface. Here, the user
supplies a separate formula for the intercepts (before: interacting) and 
for the shift (before: shifting) and avoids using the pipe \code{|} on
the left-hand-side of
the formula. Internally, the formula is translated back into the form in 
(\ref{eq:trafo}). All other functionalities in the article carry over 
to {\sc ontram}s as well. The same interface for other than ordinal 
responses is implemented in \code{dctm()}.
\begin{CodeChunk}
\begin{CodeInput}
R> dord <- data.frame(Y = ordered(sample.int(6, 100, TRUE)), 
+    X = rnorm(100), Z = rnorm(100))
R> ontram(response = ~ Y, intercept = ~ X, shift = ~ 0 + s(Z, df = 3),
+    data = dord)
\end{CodeInput}
\begin{CodeOutput}
	 Untrained ordinal outcome deep conditional transformation model


Interacting:  Y | X 

Shifting:  ~0 + s(Z, df = 3) 

Shift coefficients:
s(Z, df = 3)1 s(Z, df = 3)2 s(Z, df = 3)3 s(Z, df = 3)4 s(Z, df = 3)5 
      -0.4760       -0.7326       -0.6233       -0.4061       -0.4309 
s(Z, df = 3)6 s(Z, df = 3)7 s(Z, df = 3)8 s(Z, df = 3)9 
      -0.5447        0.6729        0.7376        0.0947 
\end{CodeOutput}
\end{CodeChunk}

\section{Large factor models}\label{app:fac}

We consider a large factor model with $10^6$ observations and a factor
variable with $10^3$ levels. The standard implementation of \code{lm()}
and \code{LmNN()} fail to process the data, due to evaluating the large
model matrix. However, we can use \code{fac_processor()} from \pkg{safareg}
to circumvent this issue and use mini-batch stochastic gradient descent
to fit the model on a standard machine. Now, \pkg{\deeptrafo} can fit
large factor models for arbitrary types of responses and censoring.
\begin{CodeChunk}
\begin{CodeInput}
R> set.seed(0)
R> library("safareg")
R> n <- 1e6
R> nlevs <- 1e3
R> X <- factor(sample.int(nlevs, n, TRUE))
R> Y <- (X == 2) - (X == 3) + rnorm(n)
R> d <- data.frame(Y = Y, X = X)
R> m <- LmNN(Y ~ 0 + fac(X), data = d, additional_processor = list(
+    fac = fac_processor))
R> fit(m, batch_size = 1e4, epochs = 20, validation_split = 0, 
+    callbacks = list(callback_early_stopping("loss", patience = 3), 
+    callback_reduce_lr_on_plateau("loss", 0.9, 2)))
R> bl <- unlist(coef(m, which = "interacting"))
R> - (unlist(coef(m))[1:5] + bl[1]) / bl[2]
\end{CodeInput}
\begin{CodeOutput}
fac(X)1 fac(X)2 fac(X)3 fac(X)4 fac(X)5 
-0.0204  0.9986 -1.0156 -0.0249  0.0477 
\end{CodeOutput}
\end{CodeChunk}
To compute the log-likelihood in models with vast amounts of data,
specifying batch-wise computation avoids memory issues.
\begin{CodeChunk}
\begin{CodeInput}
R> logLik(m, batch_size = 1e4)
\end{CodeInput}
\begin{CodeOutput}
[1] -1.42
\end{CodeOutput}
\end{CodeChunk}

\end{document}